\documentclass{aa}

\usepackage{graphicx}
\usepackage{natbib}
\usepackage{upgreek}
\usepackage{xcolor}
\usepackage{txfonts}
\usepackage{mathrsfs}
\usepackage{placeins}

\AddToHook{begindocument/before}{\usepackage{hyperref}}

\bibpunct{(}{)}{;}{a}{}{,}
\newcommand{\kms}{km\,s$^{-1}$}
\newcommand{\ecms}{ergs\,cm$^{-2}$\,s$^{-1}$}
\newcommand{\Halpha}{H$\upalpha$}

\begin{document}

%


\title{A search for Maunder-minimum candidate stars}

   \author{S.~P.~J\"arvinen\inst{1} \and K.~G.~Strassmeier\inst{1,2}}

   \institute{Leibniz-Institut f\"ur Astrophysik Potsdam (AIP),
     An der Sternwarte 16, 14482 Potsdam, Germany\\
     \email{sjarvinen@aip.de}, \email{kstrassmeier@aip.de}
     \and
     Institut f\"ur Physik und Astronomie, Universit\"at Potsdam, D-14476 Potsdam, Germany
   }

   \date{Received 12 February 2025; accepted 10 April 2025}


  \abstract
      {Stars with very low levels of magnetic activity provide an opportunity for a more quantitative  comparison with the Sun during its Maunder minimum. }
      {We employ spectra from the RAVE survey in a search for particularly low-activity stars with the goal of identifying candidates for so-called Maunder-minimum stars.}
      {Spectra were used to measure the relative flux in the cores of the \ion{Ca}{ii} infrared-triplet (IRT) lines. Those were converted to absolute emission-line fluxes and were corrected with target fluxes from high-resolution STELLA and ultra-high-resolution PEPSI spectra. }
      {Absolute \ion{Ca}{ii} IRT fluxes for a total of 78\,111 RAVE dwarf stars are presented and compared with fluxes of the 123 stars from our high-resolution STELLA+PEPSI sample. RAVE fluxes appear higher than the STELLA and PEPSI fluxes by on average 19\% for IRT-1, 21\% for IRT-2, and 25\% for IRT-3 due to their lower spectral resolution. Our sample also spans a metallicity [Fe/H] range relative to the Sun of $-1.5$ to $+0.5$\,dex. We confirm the strong  dependency of IRT fluxes on metallicity and quantify it to be at most $\pm$14\% in the B--V range 0.53--0.73. Without a metallicity correction, practically all very-low-activity RAVE dwarfs show a super-solar metallicity. After correcting for spectral resolution and for metallicity, we find 13 RAVE stars out of 13\,326 (0.1\%) that fall well below our empirical lower flux bound from high-resolution versus B--V. For solar B--V, this relates to a photospheric uncorrected radiative loss in the IRT lines of $\log R_{\rm IRT}=-4.13$ ($\approx$20\% below the solar-minimum value in late 2016). However, 11 targets turned out to be evolved stars based on their Gaia DR3 parallaxes. }
   {Two stars, TIC~352227373 (G2V) and TYC~7560-477-1 (G7V), are our only Maunder-minimum candidates from the present search. Contrary to the initial suggestion from the Mount-Wilson H\&K Survey, we conclude that such stars are very rare. }

   \keywords{stars: activity, stars: fundamental parameters, surveys, techniques: spectroscopic}

   \maketitle
%
   
\section{Introduction}

Flux-transport dynamos
\citep[e.g.][]{2008ApJ...675..920D,2012A&A...542A.127C,2021JApA...42...22H}
attempt to predict a magnetic cycle's strength and morphology. A major observational problem with the validation of a successful cycle forecast is that it takes many decades to observe the cycle if only the Sun is used as a benchmark. Furthermore, it is still not clear whether the same processes that produce cycle-to-cycle fluctuations also sometimes bring down the activity levels to produce grand minima episodes such as the Maunder minimum 
\citep[MM;][]{1894KIMS...18..173M,1904MNRAS..64..747M}.
Additional constraints may come from finding minimum chromospheric emission-line fluxes for a large sample of stars throughout the H-R diagram. Such quiet-star fluxes, together with their occurrence rate, could be seen as a part of the validation for flux-transport dynamos embedded in stellar-structure models with different effective temperatures, gravities, and metallicities. Such minimum fluxes may also serve to discover new solar analogues and are likely a universal phenomenon
\citep[][and references therein]{2012A&A...540A.130S}.

The presence of emission in the cores of the \ion{Ca}{ii}~H\&K and infrared-triplet (IRT) resonance lines, the latter at $\lambda$8498 (dubbed IRT-1), $\lambda$8542 (IRT-2), and $\lambda$8662 (IRT-3)\,\AA, is a simple diagnostic of magnetic activity in the chromospheres of late-type stars 
\citep[e.g.][]{1979ApJS...41...47L,1979ApJS...41..481L,1993ApJS...86..293D,2000A&A...353..666C,2005A&A...430..669A,2017A&A...605A.113M,2018A&A...616A.108B,2023A&A...674A..30L}.
Spatially resolved \ion{Ca}{ii} heliograms and magnetograms have amply demonstrated the relation between H\&K-emission strength and the surface magnetic field on our Sun
\citep[][a.o.]{1972SoPh...24...98F, 1996IAUS..176....1S}.
Furthermore, the fact that we observe generally stronger \ion{Ca}{ii} emission in more rapidly rotating stars is widely known as the rotation-activity relation
\citep[e.g.][]{1984ApJ...287..769N},
which is heuristically explained by the $\Omega$ effect of the classical $\upalpha\Omega$ dynamo
\citep[e.g.][]{2003SoPh..212....3S}.
Anomalies such as grand activity minima are important constraints for such dynamo theories.

The Mount Wilson Observatory (MWO) H\&K survey revealed a significant number of solar-type stars with chromospheric emission-line fluxes smaller than even the Sun's in its activity minimum but without cyclic variations
\citep[so called ``flat'' stars;][]{1995ApJ...438..269B}.
Some of these stars were interpreted as being in a MM-like state and numerous questions were raised, such as how their coronal and chromospheric heating processes would function. However, further studies have shown that almost all of these flat stars were in fact older sub-giants
\citep{2004ApJS..152..261W}
with diminished surface fluxes, and therefore not MM candidates.
\citet{2013A&A...554A..50S}
investigated the Mount Wilson survey stars further and took the effects of metallicity into account also. They concluded that the stars in the different activity groups populate different parts of the main-sequence evolution. 

Our own Sun has been telescopically observed now for more than 400 years 
\citep[e.g.][]{1976Sci...192.1189E},
while our records for some of the MWO survey stars have reached an impressive five decades of S-index measurements 
\citep[e.g.][]{2017ApJ...835...25E,2022AJ....163..183B}.
Ongoing monitoring of selected stars remains the most direct way of searching for Maunder minima. 
\citet{2018ApJ...863L..26S} 
studied the activity cycle of HD\,4915 and found a pattern suggesting the beginning of a magnetic grand minimum. A similar pattern was seen by   
\citet{2022AJ....163..183B} and 
\citet{2022ApJ...936L..23L},
who identified one more MM candidate, HD\,166620, out of a sample of 59 MWO stars, which transitioned from cycling to flat MM-like activity. 
\citet{2024ApJS..274...35I}
published chromospheric activity cycles for 138 main-sequence stars from the Keck/HIRES California Legacy Survey and found 50 stars that had a smaller S-value standard deviation than HD\,10700 ($\tau$\,Ceti), indicating very low activity levels. Despite the fact that only long-term monitoring is eventually able to prove the existence of a MM, we still need more candidate targets. 

Just recently, 
\citet{2023A&A...674A..30L}
presented a large subset of the {\it Gaia} DR3 spectra of spectral resolution $R$=11,500. They measured \ion{Ca}{ii} IRT radiative losses for two million stars. The interest in the dynamical evolution of our Galaxy has also spurred large-scale ground-based stellar surveys, most of which use the \ion{Ca}{ii} IRT as a tracer for radial velocity and chemical abundance. Such a survey at a comparable low resolution ($R$=7,500) is the already completed Radial Velocity Experiment
\citep[RAVE;][]{2006AJ....132.1645S}.
For the present paper, we selected the late-type dwarfs from this database and measured the \ion{Ca}{ii} IRT line-core flux in $\approx$80,000 spectra. A catalogue of chromospherically active stars had already been constructed
\citep{2013ApJ...776..127Z}
and partly analysed
\citep{2017ApJ...835...61Z}
based on a subset of RAVE data. To reliably extent the measurement range towards the lowest possible fluxes, we have also obtained high-resolution follow-up spectra of selected joint RAVE targets as well as spectra of Morgan-Keenan (M-K) standard (dwarf) stars using the \'echelle spectrograph of the STELLar Activity (STELLA) telescope. Superior spectra of solar-like quality of a sub-sample of the bright {\it Gaia} benchmark stars from the Potsdam Echelle Polarimetric and Spectroscopic Instrument (PEPSI) at the 11.8m Large Binocular Telescope (LBT) are now used in the present paper to derive \ion{Ca}{ii} IRT fluxes of well-characterised inactive stars including the Sun-as-a-star. These spectra cover all wavelengths from 390--900\,nm and can therefore relate IRT fluxes to those of other activity indicators, such as \ion{Ca}{ii} H\&K and \Halpha.
 
We describe the new observations and the three different instrument samples, RAVE, STELLA, and PEPSI, in Sect.~\ref{sect:obs}. The methods and techniques for converting the relative intensity spectra to absolute emission line fluxes, along with their shortcomings and corrections, are described in Sect.~\ref{Snew}. Section~\ref{S3} derives the IRT fluxes and applies a spectral-resolution correction. Section~\ref{analysis} is the core of our analysis to identify MM candidates. Our conclusions are drawn in Sect.~\ref{sec:conc}.  

\section{Observations and sample definitions}\label{sect:obs}

\subsection{RAVE low-resolution spectroscopy}

\begin{figure}
\centering
  \includegraphics[width=\columnwidth]{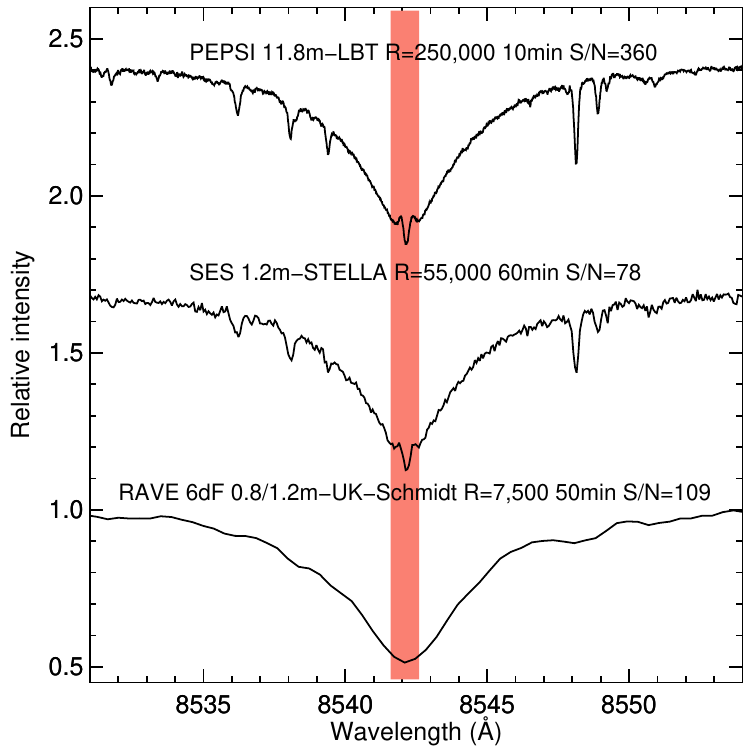}
  \caption{
  \ion{Ca}{ii} IRT-2 line profile of HD\,82106 (K3\,V, $v\sin i = 3.0$\,\kms) at three different spectral resolutions: RAVE $R$=7,500; STELLA $R$=55,000; and PEPSI $R$=250,000. The spectra are on the same scale but shifted in intensity for better viewing. The vertical shaded area indicates the 1-\AA{} integration window for the line-core fluxes.
  }
  \label{F-pepsi}
\end{figure}

Observations were obtained using the 1.2-m UK Schmidt Telescope (UKST) of the Australian Astronomical Observatory (AAO) between 2003 and 2013. The spectrograph itself was known as 6dF
\citep{2000SPIE.4008..123W}
in reference to its 6\degr\ diameter field of view. The facility was decommissioned after the survey ended. The survey has produced in total six data releases. The final data release
\citep[DR6;][]{2020AJ....160...83S, 2020AJ....160...82S}
has 518,387 $R = \lambda/\Delta\lambda \approx 7,500$ (0.11\,nm, or 38\,\kms{}) spectra of 451,783 stars covering the \ion{Ca}{ii} IRT. Sampling of one resolution element is with 3.1 pixels. The typical signal-to-noise ratio (S/N) is 50 per pixel, with the majority of targets within a $\pm$25 range. A well-exposed example spectrum is shown in Fig.~\ref{F-pepsi}. 

\subsection{STELLA high-resolution spectroscopy}

New observations of selected (northern) RAVE targets as well as a number of M-K standard stars were obtained with the STELLA Echelle Spectrograph
\citep[SES;][]{2012SPIE.8451E..0KW}
fed by the robotic 1.2-m STELLA-II telescope at the Observatorio del Teide in Tenerife, Spain
\citep{2004AN....325..527S, 2010AdAst2010E..19S}.
These spectra cover the wavelength range from 386.5~nm to 882.5~nm with a 3-pixel resolving power of $R$=55,000, corresponding to a spectral resolution of 0.015\,nm (5.2~\kms) at 860\,nm. The S/N per pixel varies significantly within a single exposure due to the large wavelength coverage. The blue part of the spectrum is usually the least exposed part. The typical peak S/N for an eighth-magnitude star is 100:1 in one hour in the red, and 10:1 at the blue cut-off. STELLA-SES spectra were automatically reduced and extracted using the IRAF-based data-reduction pipeline
\citep[see][]{2012SPIE.8451E..0KW}.
Details of the data reduction and analysis of STELLA spectra can be found, for example, in
\citet{2012AN....333..663S}.

\subsection{PEPSI ultra-high-resolution spectroscopy}

Ultra-high-resolution spectra with a two-pixel resolution of up to 250,000 (corresponding on average to 0.004\,nm or 1.36~\kms\ at 860\,nm) were obtained for 48 {\it Gaia} benchmark stars 
\citep[GBS;][]{2014A&A...566A..98B}
and the Sun-as-a-star as presented earlier by 
\citet{2018A&A...612A..45S}.
These spectra were obtained employing PEPSI
\citep{2015AN....336..324S}
at the 2$\times$8.4\,m LBT in Arizona and they cover the wavelength range from 383 to 912\,nm without gaps. We now use these spectra for an analysis of their \ion{Ca}{ii} IRT lines. The S/N in the \ion{Ca}{ii} IRT wavelength regime is between 200 and 4300 per pixel, depending on the brightness of the target and the total number of available exposures. One of these GBS targets was also observed in RAVE and by STELLA (\object{HD 82106}, K3\,V)\footnote{This spectrum and all other spectra of northern-hemisphere GBS targets were published as a spectral atlas as part of a deep-spectrum survey
\citep{2018A&A...612A..45S};
they are downloadable at \url{https://pepsi.aip.de}}. It is among the lower-S/N targets in the PEPSI GBS sample, peaking at 360 per pixel at 860\,nm, which, however, is still 3--4 times higher than for STELLA and RAVE. Two back-to-back exposures of 5\,min each were co-added to reach this S/N in the continuum. Figure~\ref{F-pepsi} is a comparison of its IRT-2 (8542\,\AA ) line profile with STELLA and RAVE. 

\subsection{Late-type dwarfs in the RAVE sample}\label{sect:dwarfs}

Our initial selection of the late-type dwarfs from the full RAVE sample was based on the effective temperature of $T_{\rm eff} \leq 7500$~K and a logarithmic gravity of $\log g \geq 4.0$, determined in DR6. Additionally, only targets for which the quality flag was `0' (i.e.\ the analysis was carried out as desired) were investigated. These selection criteria resulted in 86,911 individual spectra. In a second round of selection, some of the initial spectra had to be rejected due to either a too low S/N (6,648 spectra,  $\sim$7.6\%) or having the Ca triplet feature blended by strong and broad Paschen lines (1,094 spectra; $\sim$1.3\%). In addition to that, 862 spectra (<1\%) were rejected due to bad continuum normalisation, problems with the wavelength calibration, or binarity and multiplicity. Usually the same spectrum could have been rejected for several reasons. As we suspected, a number of these stars ended up in later inspections likely not being dwarf stars because the low S/N could not constrain the stellar parameters well enough. Examples of such spectra of targets that had to be rejected are shown in the Appendix in Fig.~\ref{AFi:bad}. Our initial working sample was then confined to S/N$>$15 spectra and consisted of 78,111 bona fide dwarf stars.

The stellar parameters we are interested in, namely the effective temperature, gravity, and metallicity, released in DR6, are based on the analyses already presented in DR4
\citep{2013AJ....146..134K}
and DR5
\citep{2017AJ....153...75K}.
In DR5, the stellar parameters were computed using the RAVE DR4 stellar pipeline, but a new calibration increased the accuracy of the stellar parameters by up to 15\%. This calibration was employed mainly because at that time there were RAVE stars with $\log g$ values determined asteroseismically
\citep{2017A&A...600A..66V}.
Furthermore, the metal-rich tail of the RAVE stars was also re-investigated, by increasing the number of calibration stars in the supersolar metallicity regime. However, the updated DR5 stellar parameters mainly improved the gravities of the giants and the stars with supersolar metallicities.
\citet{2014AJ....148...81M}
also showed that DR4 effective temperatures for warm stars ($T_{\rm eff} \ga$ 6000~K) are underestimated by about 250\,K (for details, we refer to DR6). As is described in DR6,  the metallicity should be thought of as a metallicity indicator and [Fe/H]$_\mathrm{DR6}$ is possibly not equal to the overall metallicity of the star. The errors for these quantities have already been summarised by
\citet{2013AJ....146..134K},
and are typically 170~K for temperature, 0.27~dex for $\log g$, and 0.16~dex for metallicity.

\subsection{STELLA and PEPSI samples}

We recall that the RAVE targets were observed from the southern hemisphere, while the STELLA and LBT observatories are located in the northern hemisphere. This limits the number of joint targets. Therefore, only targets with declination $> -30$\degr{} could be considered but, furthermore, we required that the target's spectral classification and metallicity be known with sufficient precision. A secondary selection criterion had been the visual brightness of the target in order to keep the exposure times at STELLA reasonable (because STELLA is just a 1.2\,m telescope but its spectrograph delivers $R$=55,000 sampled by 3 CCD pixels). The number of joint RAVE and STELLA targets is thus just 23, listed in the Appendix in Table~\ref{BT:Targets}, but these evenly cover the F0--K6 spectral range. In addition, we have observed 49 northern M-K spectrum standard stars with STELLA, covering spectral types from F0 to M2
\citep{1989BICDS..36...27G}.
These are also listed in the appendix in Table~\ref{BT:MKs}.

For the one joint target (HD\,82106, Table~\ref{T:pepsi}), we present fluxes at three different spectral resolutions. Although the observations were taken at different times, it appears that HD\,82106 is a normal (inactive) K3 dwarf, and dramatic chromospheric changes in its line core over time are not expected. However, the star has never been monitored, nor do we know anything about the presence of rotational modulation.

\begin{table}
  \caption{
  Absolute \ion{Ca}{ii} IRT emission line fluxes, $\mathscr{F}_\mathrm{IRT}$, for HD\,82106 in $10^5$ \ecms.
  }
\centering
\label{T:pepsi}
\begin{tabular}{lccc}
\hline\hline\noalign{\smallskip}
                    & RAVE           & STELLA          & PEPSI \\
IRT-1 $\lambda$8498 & $15.8 \pm 2.4$ & $14.5 \pm 2.1$ & $14.7 \pm 2.1$ \\
IRT-2 $\lambda$8542 & $12.5 \pm 1.9$ & $12.0 \pm 1.8$ & $12.3 \pm 1.8$ \\
IRT-3 $\lambda$8662 & $12.8 \pm 2.0$ & $11.2 \pm 1.7$ & $11.8 \pm 1.7$ \\
\hline
\end{tabular}
\end{table}

\begin{figure}
\centering
  \includegraphics[width=\columnwidth]{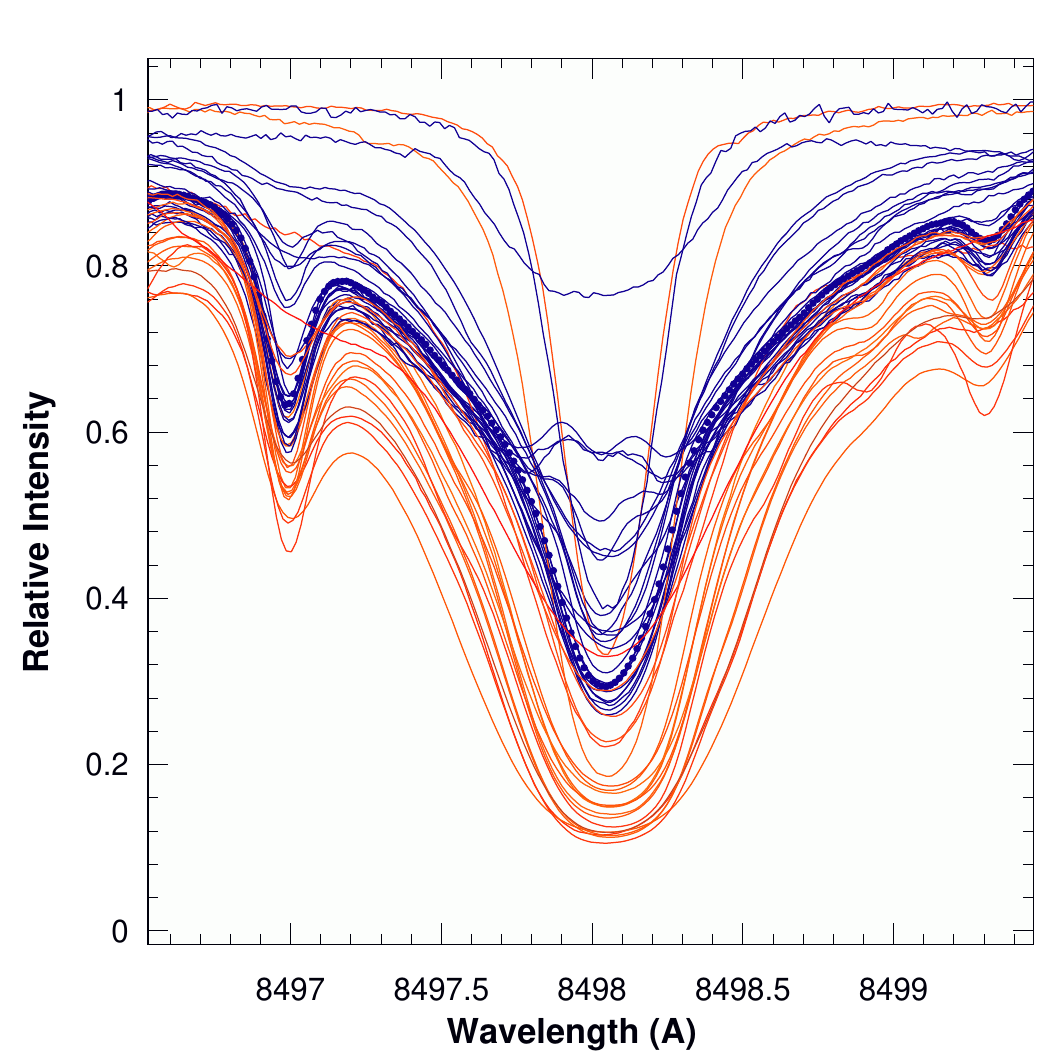}
  \caption{
  Ultra-high-resolution \ion{Ca}{ii} IRT-1 line profiles of the {\it Gaia} benchmark stars observed with PEPSI (IRT-2 and IRT-3 are shown in the appendix in Fig.~\ref{F_gbs2-3}). Blue spectra are the dwarf stars, red spectra the giants. The solar spectrum is shown as dots. Note that the width of the graph of 3\,\AA{} equals the flux integration window employed by 
  \citet{2023A&A...674A..30L}
  for the {\it Gaia} DR3 IRT fluxes.
  }
  \label{F-gbscomp}
\end{figure}

The PEPSI sample is made of the northern-hemisphere {\it Gaia} benchmark stars 
\citep{2014A&A...566A..98B}
visible from the site of the LBT in Arizona. There is a total of 48 such inactive AFGKM stars (including the Sun but omitting HD\,189333), 26 dwarfs and 22 giants, which have a very high-S/N $R$$\approx$250,000 PEPSI spectrum 
\citep{2018A&A...612A..45S}.
Table~\ref{BT:gbs} in the Appendix lists these target along with $v\sin i$, $T_{\rm eff}$, $\log g$, and [Fe/H] updated whenever available from the most recent compilation of 
\citet{2024A&A...682A.145S}. 
The respective $(B-V)_0$ values were determined from the spectroscopic $T_{\rm eff}$ and the $(B-V)_0$-$T_{\rm eff}$ calibration from 
\citet{2000AJ....120.1072S}.
It takes into account gravity and metallicity and succeeds the previous conversion from 
\citet{1996ApJ...469..355F}.
The value for $(B-V)_0$ of the Sun from $T_{\rm eff}$=5770\,K is +0.629, in good agreement with literature values from 
\citet{1971MNRAS.155...65B},
\citet{1971PASP...83...79F},
or
\citet{1972PASP...84..515C},
which we adopt in this paper. Figure~\ref{F-gbscomp} is an over-plot of the IRT-1 line profiles of these 48 benchmark targets (IRT-2 and IRT-3 are shown in the appendix in Fig.~\ref{F_gbs2-3}) and illustrates the large range of line morphology if one compares A-to-M stars, giants, and dwarfs. 

\section{Calcium IRT line-core fluxes}\label{Snew}

\subsection{Calibration and definition of absolute fluxes}\label{S2d}

In this paper, we integrate the central 1-\AA\, portion of each \ion{Ca}{ii} IRT line profile based on continuum-normalised spectra. This relative flux, $f_\mathrm{IRT}$, is then multiplied with the absolute continuum flux, $\mathscr{F}_{\mathrm c}$, 
\begin{equation}
  \mathscr{F}_\mathrm{IRT} = f_\mathrm{IRT} \ \mathscr{F}_{\mathrm c} \ ,
\end{equation}
obtained from the relations provided by
\citet{1996PASP..108..313H}
for various M-K luminosity classes and colour ranges. The relations are
\begin{equation}
  \log \mathscr{F}_{\mathrm c} = 7.223 - 1.330\, (B - V)
\end{equation}
for M-K~I-V and --0.1 < (B--V) < 0.22, and
\begin{equation}
  \log \mathscr{F}_{\mathrm c} = 7.083 - 0.685\, (B - V)
\end{equation}
for M-K~I-V and 0.22 < (B--V) < 1.4. The original relations were obtained from calibrations with stars in a temperature range of $\approx$4100--7800~K, meaning that the coolest stars -- all M dwarfs -- are outside of its validated range. However, extrapolating $\mathscr{F}_{\mathrm c}$ to B--V of 1\fm5 ($T_{\rm eff}\approx 3800$~K) will not severely jeopardise the validation, but just increase the root mean square (rms) uncertainty, and then will include  basically all of our coolest RAVE targets. An extended relation including M dwarfs as red as B--V of 1\fm8 was given by
\citet{2007A&A...469..309C}
for the 3950-\AA\, continuum, which we also take as a principal validation for the \ion{Ca}{ii} IRT wavelength range.

For the RAVE spectra, we have only 3.6 pixels for the 1-\AA{} window, whereas in the case of the STELLA spectra we have 13.3 pixels for the same 1-\AA{} window, and in the case of PEPSI the number is 63 pixels. Thus, before we integrate the central 1-\AA{} portion of each RAVE IRT line profile, its spectrum is interpolated into 0.1\,\AA{} bins to minimise pixelised windowing effects. 

In this way, the relative flux (and equivalent width) increases with increased line-core filling due to chromospheric emission, while the continuum flux reflects the dependence on stellar radius and $T_{\rm eff}$. While simple and easy to measure, the extracted line-core flux depends on the initial continuum setting of the spectra. Its errors map linearly into the fluxes. However, this is more of a problem for the H\&K wavelength region, where there is basically no continuum visible and continuum-setting errors may amount to up to 20\% or even higher. It is still a concern for the IRT wavelength region for  \'echelle spectra, where the resonance-line wings may span a full \'echelle order or even extend into different orders. Then, continuum errors of up to a few percent may be possible (see next section), but otherwise the continuum setting at IRT wavelengths is less critical than at H\&K (but see 
\citet{2000A&A...353..666C} 
for the impact of the hydrogen Paschen lines for the hotter stars).

Part of the 1-\AA\ line-core flux will be photospheric in origin. Its removal can be tricky since, for example, empirical photospheric corrections with bona fide inactive stars may not be only photospheric in origin, or these stars' photospheric structures are possibly very different, or the metallicities simply do not match   
\citep[see, e.g.][]{1985ApJ...294..626C, 1993ApJS...85..315S, 2007A&A...469..309C, 2005A&A...430..669A}.
The correction is particularly difficult for the \ion{Ca}{ii} IRT lines because of the even lower photosphere-chromosphere contrast when compared to H\&K. Purely synthetic corrections based on a radiative equilibrium model with an adopted $T_{\rm eff}$ 
\citep[see][]{1979ApJS...41...47L}  
introduce flux-correction errors via $T_{\rm eff}^4$. Because a photospheric correction is not relevant to a search for MM candidates, we decided to apply no photosphere correction at all. 

More detailed LTE calculations by 
\citet{2000A&A...353..666C} 
showed a fairly strong dependency of the IRT wings and line-core depth on metallicity in the sense that lower metallicities cause shallower lines, as for the optically thin lines. The sensitivity to metallicity is generally higher for the hotter stars.  
\citet{2005A&A...430..669A} 
and 
\citet{2007A&A...461..261M}
concluded that NLTE effects strengthen the \ion{Ca}{ii} lines and lead to enhanced absorption in the line core. Although NLTE corrections remain small, $<$0.02\,dex, they will increase with decreasing abundance and metallicity. The Phoenix NLTE IRT profiles presented in 
\cite{2017A&A...605A.113M} 
even show an inverse behaviour in the core compared to the wings: lower metallicity causes deeper line cores instead of shallower ones. Dynamic 3D abundance effects 
\citep[e.g.][]{2024arXiv240100697L} 
have yet to be explored.

As can be seen in Fig.~\ref{F-pepsi}, the narrow 1-\AA\ central bandpass already minimises the photospheric contribution because it appears that the width of the chromospheric self reversal (from the blue minimum point to the red minimum point) is $\approx$1\,\AA\ for the majority of stars in our sample. Although 
\citet{2017A&A...605A.113M} 
provide a convenient collection of such correction fluxes based on 26 bona fide inactive stars as a function of B--V and $v\sin i$, the numerical values are between 4$\times$10$^6$ and 7$\times$10$^6$\,\ecms\ for the sum of all three IRT lines for B--V of 0.8 to 0.5, respectively. These corrections  are usually only marginally smaller (if at all) than the measured total fluxes for our MM candidates, and thus problematic to apply. This is yet another reason why we  refrained from a photospheric correction, synthetic or with bona fide inactive stars, and give combined photospheric and chromospheric 1-\AA\ IRT fluxes throughout this paper. Again, a photospheric correction is not relevant to our search for MM candidates as long as we compare stars within small B--V bins (see later in Sect.~\ref{analysis}).  

In any case, the largest contribution to the (absolute flux) error is due to the continuum-flux calibration and comes mostly from the error in $T_{\rm eff}$ but also from systematic differences in the $T_{\rm eff}$-colour relation. $\mathscr{F}_{\mathrm c}$ is based on B--V from the conversion of the spectroscopically determined $T_{\rm eff}$, $\log g$, and metallicity
\citep[for RAVE taken from][]{2020AJ....160...82S} 
with the colour transformation from
\citet{2000AJ....120.1072S}. 
All B--V values in this paper are based on this transformation. 

We also note that the basal flux level -- the chromospheric emission unrelated to magnetic activity -- remains included in our flux measurements. Such a basal flux is present in even the most inactive stars and acts like a B--V-dependent offset for the flux. According to the study by
\citet{Martinphd} (see also 
\citet{2017A&A...605A.113M}),
the IRT lines always show a lower basal flux level than the \ion{Ca}{ii} H\&K lines, ranging from roughly 85\% of the basal flux at the bluest B--V, where the relation may be least trustworthy, to about half of its flux level at $0.4 < B-V < 0.8$. Fortunately, neither of these corrections, the photospheric contribution or the basal flux, are decisive for the main aim in this paper of finding MM candidates. 

\subsection{Continuum setting}\label{cont}

We employed our joint RAVE-STELLA target sample, together with the high-quality GBS PEPSI spectra, to verify the continuum setting around the IRT lines. While RAVE and STELLA employ IRAF subroutines and standard sigma clipping following either bi-cubic spline or polynomial fits, the PEPSI data reduction 
\citep[SDS4PEPSI;][]{2023AN....34430059I}
is far more elaborate and based on predetermined and tabulated synthetic spectra
\citep{2018A&A...612A..44S}. 
SDS4PEPSI performs a global continuum fit across the full CCD coverage, while the IRAF-based SES pipeline fits a continuum to each \'echelle order separately. The PEPSI continuum is usually defined to a precision of 0.1\,\%\ or even better depending on the S/N. In the case of RAVE spectra, the continuum is normalised using an iterative low-order polynomial fitting with asymmetric rejection limits
\citep{2006AJ....132.1645S}.

\begin{figure}
\centering
  \includegraphics[width=\columnwidth]{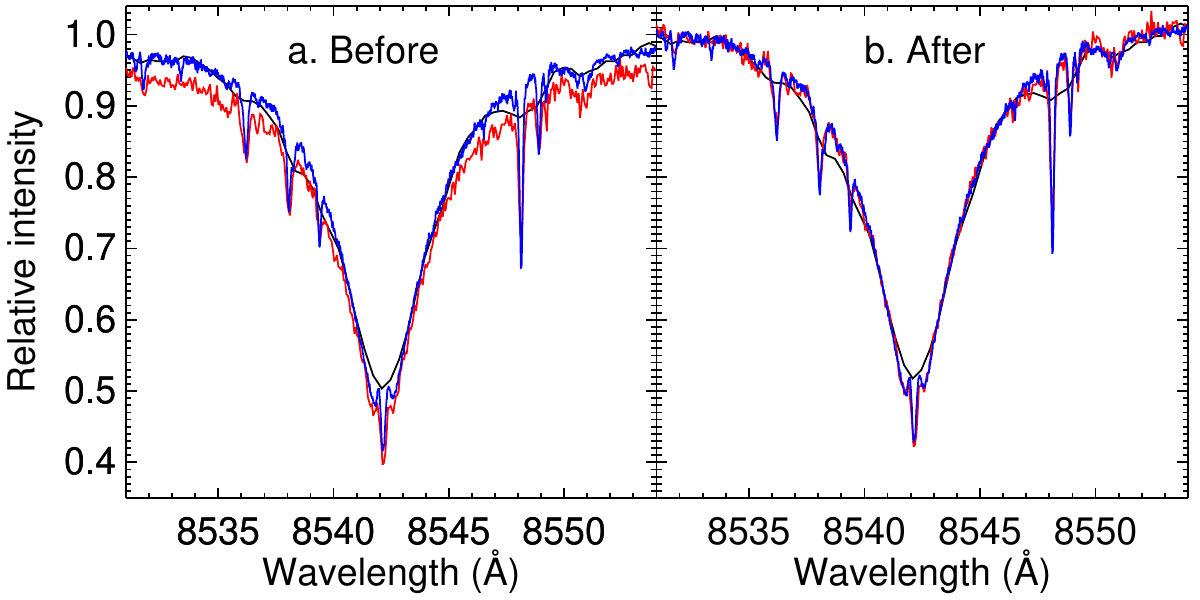}
  \caption{
  \ion{Ca}{ii} IRT-2 line profile of HD\,82106 visualising the initially different continuum levels with RAVE (black), STELLA (red), and PEPSI (blue). 
    {\bf a.} Before continuum re-normalisation.
    {\bf b.} After re-normalisation to the PEPSI continuum.
    }
  \label{F-pepsicomp}
\end{figure}

When we match the RAVE continuum for HD\,82106 to the PEPSI continuum, we find no systematic offset in the line wings of the IRT-1 and IRT-2 lines (see Fig.~\ref{F-pepsicomp}a), but there is a 1\%\ offset for the IRT-3 line. The same procedure shows a 0.3\%\ offset for the STELLA spectra in the IRT-1 line but a 3.1\%\ offset in the IRT-2 line, and a 1.5\%\ offset in the IRT-3 line. These relatively large offsets for STELLA spectra are due to the imperfect removal of the blaze function for the orders that contain broad resonance lines such as the IRT
\citep[see][]{2008SPIE.7019E..0LW}.
The STELLA spectra of HD\,82106 were re-normalised for each IRT line region before flux determination. The outcome of this re-normalisation is illustrated in Fig.~\ref{F-pepsicomp}b, labelled `After'. A similar re-normalisation was then applied to all STELLA spectra in this paper.

\subsection{Rotational line broadening}\label{broad}

Since accurate $v\sin i$ values are difficult to measure at $R=7500$, the effect of rotational line broadening on RAVE emission line fluxes remains uncertain or must even be ignored. As was discussed in
\citet{2012AN....333..663S},
the corrections are negligible for stars with $v\sin i < 10$~\kms{}, but would be 30\%\ for a rapidly rotating (cool) star with $v\sin i = 60$~\kms. However, rapidly rotating (cool) stars are comparably rare and are always higher-flux stars due to the rotation-activity relation, or are binaries, and thus are not candidates for a MM state anyway. Our STELLA calibration sample (on purpose) does not include rapidly rotating cool stars. All comparably rapidly rotating stars are F dwarfs with a B--V colour bluer than that for the Sun. Fluxes for the cases with $v\sin i$ greater or equal to 10.0\,\kms\ (see Fig.~\ref{Fi:SESfluxvscol} in the Appendix) are corrected with the appropriate value provided by Eq.~(5) in
\citet{2012AN....333..663S}.
For this purpose, we redetermined $v\sin i$ from the STELLA spectra by applying our tool ParSES; we list these values in the tables in the Appendix. ParSES
\citep[see, e.g.][]{2013POBeo..92..169J}
fits a synthetic spectrum from a pre-tabulated library to five selected \'echelle orders of the observed spectrum. Radial-tangential macroturbulence was removed based on the values tabulated in
\citet{2005oasp.book.....G}.
The average $v\sin i$ errors are 1~\kms{} for values below 10--15~\kms{}, 1--2~\kms{} for values between 15--25~\kms{}, and 2~\kms{} for values greater than that. 

\subsection{Line blending}\label{blend}

The large wavelength coverage of the three IRT resonance lines naturally contains many line blends from other elements. The most severe blends are from the hydrogen Paschen series for the hotter stars and the myriads of (mostly) CN molecular lines for the coolest stars. Most of these blends are bypassed by the narrow 1-\AA\ line core window. However, 
\citet{2023A&A...674A..30L}
employed a 3-\AA\ wide integration interval for the {\it Gaia} DR3 data release instead of the usually employed 1-\AA\ for ground-based spectra. This is not critical for extracting a chromospheric flux but is critical when one wants to compare stars and look at the $T_{\rm eff}$ dependence because it includes several line blends from other elements for which the line strengths likely change with photospheric temperature (and metallicity) differently than for singly ionised Ca. For example, \ion{Ca}{ii} IRT-3 has a strong iron blend close to the line core (Fe\,{\sc i} 8661.9\,\AA ),\footnote{The sunspot umbral spectrum 
\citep[][$T_{\rm umbra} \approx 4000$\,K]{1999asus.book.....W}
shows this blend actually as a resolved Zeeman triplet.} while \ion{Ca}{ii} IRT-1 has at least two blends, the stronger one at 8497.0\,\AA\ (also \ion{Fe}{i}). These blends are not within the 1-\AA\ integration window, or are at least minimised then, but are all in the 3-\AA\ window applied for the {\it Gaia} DR3 data release. We thus caution against using it, in particular when comparing minimum fluxes of bona fide inactive stars.

\section{Towards minimum IRT fluxes}\label{S3}

\subsection{Application of flux corrections}\label{zeropointcorr}

Our temporary IRT fluxes were consolidated after the removal of the effects from continuum-setting systematics and rotational line broadening (whenever known). No blending corrections were applied. We recall that the continuum-setting issue was addressed in Sect.~\ref{cont} and applied  only to the STELLA-SES spectra. RAVE and PEPSI spectra were not affected. 

Rotational line broadening for the RAVE targets was unknown, and thus not corrected for, while the respective $v\sin i$ measures from STELLA and PEPSI spectra were used for their correction (listed in the tables in the Appendix). This inconsistency is not critical, though, because the low spectral resolution of RAVE makes the spectra insensitive to the line-broadening contribution for all but the fastest rotators. Therefore, a correction for the slowly rotating cool stars under consideration for MM appears negligible.  

\begin{figure}
  \centering
  \includegraphics[width=\columnwidth]{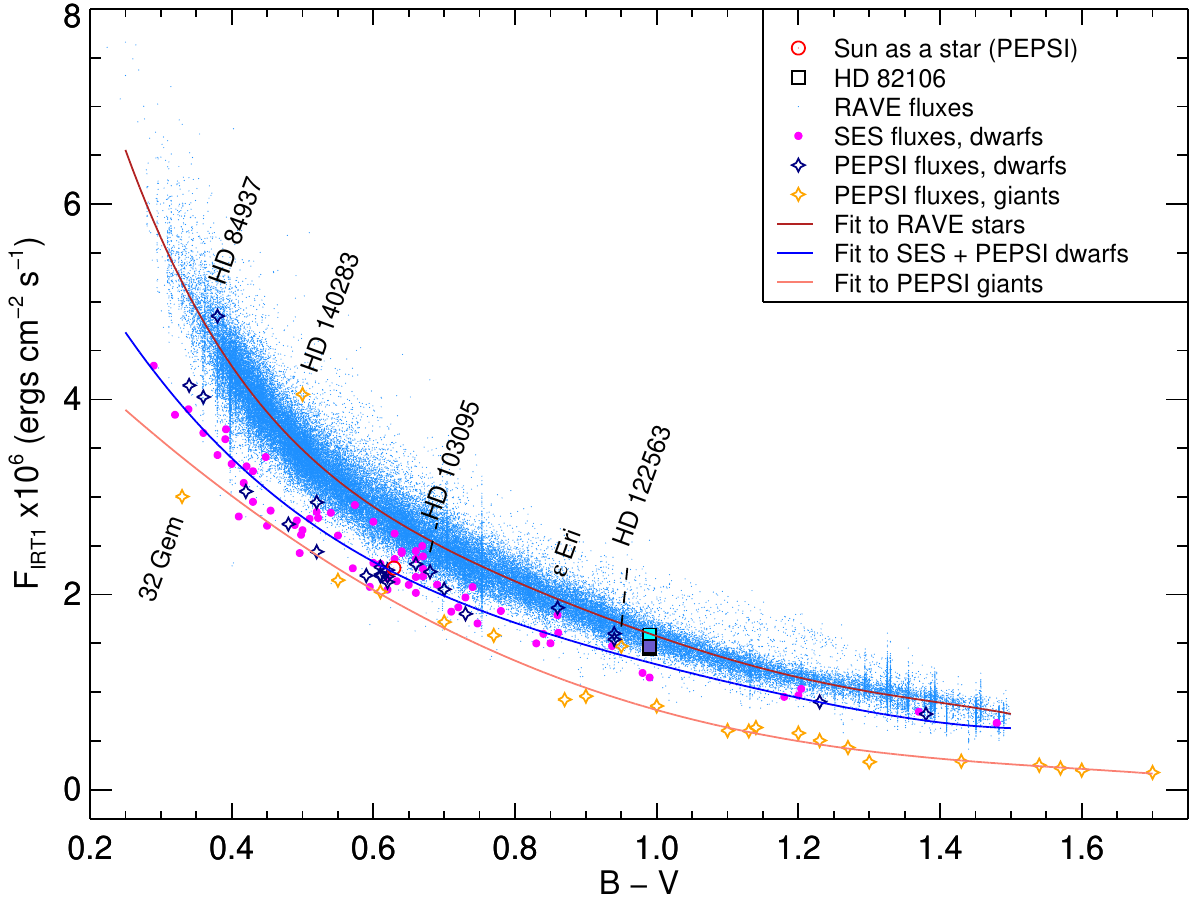}
  \caption{
  \ion{Ca}{ii} IRT-1 fluxes for our three instrument samples. RAVE data points are plotted with light blue dots, except for HD\,82106 for which a cyan square is used. The fit to them is presented with a dark red line. STELLA-SES data points, both the common sample with RAVE as well as the MK standards are plotted with filled magenta circles, HD\,82106 is plotted with a light blue square. PEPSI data points are plotted as follows: blue diamonds are used for dwarfs, yellow diamonds are used for giants, the Sun is plotted with red solar symbol, and HD\,82106 is plotted with a violet square. The orange fit is for the giants only. The blue fit is based on both STELLA-SES and PEPSI dwarfs. RAVE fluxes appear on average higher than STELLA-SES and PEPSI fluxes by 19\% for IRT-1, 21\% for IRT-2, and 25\% for IRT-3. 
  }
  \label{Fallsamples}
\end{figure}

With the above flux corrections applied, we now assume that the average flux per B--V bin represents the normal photospheric-plus-chromospheric IRT flux expected for an inactive or only mildly active star. Stars with fluxes significantly below these minimum fluxes are, then, our subject of interest for Maunder minima. The average fluxes as a function of B--V for each sub-sample -- RAVE, STELLA, and PEPSI -- are defined by a polynomial fit as obtained and described in the following subsections and shown in Fig.~\ref{Fallsamples}.  

\subsection{Low-resolution RAVE fluxes}

The total of 78,111 targets also fulfilled the absolute flux calibration limits by
\citet{1996PASP..108..313H}
extended to B--V=1.5 on the red side. All of these spectra were measured in the way described in Sect.~\ref{S2d} for the three IRT lines. A linear presentation of the absolute \ion{Ca}{ii} IRT line-core fluxes for IRT-1 is plotted against the B--V colour index in Fig.~\ref{Fallsamples} (IRT-2 and IRT-3 are shown in the appendix in Fig.~\ref{Fall_IRT2-3}). Because there are not homogeneously measured B--V colours for all the stars, we applied the transformation by 
\citet{2000AJ....120.1072S}
and converted the RAVE temperatures along with gravities and metallicities (from DR6) back to B--V colours. 

The obtained average fluxes are described using five orders for the B--V colour term as follows:
\begin{eqnarray}\label{fit1}
  \mathscr{F}_{\mathrm{IRT-1}}^{\mathrm{RAVE}} & = & (14.87 -51.19(B - V) + 90.17B - V)^{2} \nonumber\\
  & & -84.80(B - V)^{3} +39.87(B - V)^{4} \nonumber\\
  & & -7.35(B -V)^{5})\times 10^{6} \ ,
\end{eqnarray}
\begin{eqnarray}\label{fit2}
  \mathscr{F}_{\mathrm{IRT-2}}^{\mathrm{RAVE}} & = & (13.11 -51.45(B - V) + 98.27(B - V)^{2} \nonumber\\
  & & -96.63(B - V)^{3} +46.61(B - V)^{4} \nonumber\\
  & & -8.72(B -V)^{5})\times 10^{6} \ ,
\end{eqnarray}
\begin{eqnarray}\label{fit3}
  \mathscr{F}_{\mathrm{IRT-3}}^{\mathrm{RAVE}} & = & (13.90 -53.81(B - V) + 102.62(B - V)^{2} \nonumber\\
  & & -101.95(B - V)^{3} +50.06(B - V)^{4} \nonumber\\
  & &  -9.58(B - V)^{5})\times 10^{6} .
\end{eqnarray}
Figures~\ref{Fallsamples} and \ref{Fall_IRT2-3} also show the unweighted fits according to the equations in Eq.~\ref{fit1}--\ref{fit3}, respectively. The scatter in flux values is smallest for the IRT-1 line ($\lambda$8498), 21\%, whereas the IRT-2 ($\lambda$8542) and IRT-3 ($\lambda$8662) flux values show larger scatters, 23\% and 24\%, respectively. 

\subsection{High-resolution STELLA fluxes}

The IRT fluxes from the STELLA spectra were obtained the same way as for the RAVE spectra and are also shown for IRT-1 in Fig.~\ref{Fallsamples} (for IRT-2 and IRT-3, see the appendix, Fig~\ref{Fall_IRT2-3}). Our main intention was to calibrate the zero point of the RAVE fluxes to quantify the influence of spectral resolution. This is explicitly shown for the joint sample of 23 targets in the Appendix in the log-log Fig.~\ref{Fi:fluxSR}. The scatter is comparable for the three IRT lines, and the (log) rms is 0.08 for IRT-1 and for IRT-2, and 0.10 for IRT-3. Sampling becomes sparser at the lower flux end but the slope is reproduced and consistent for all three IRT lines. As was expected, RAVE $R$=7,500 fluxes appear consistently larger than the STELLA $R$=55,000 fluxes on average by 19\% for IRT-1, 21\% for IRT-2, and 25\% for IRT-3.

The obtained average fluxes are sufficiently well described using four orders for the B--V colour term:
\begin{eqnarray}\label{fit1R}
  \mathscr{F}_{\mathrm{IRT-1}}^{\mathrm{STELLA}} & = & (8.10 -18.56(B - V) + 21.74(B - V)^{2} \nonumber\\
  & & -13.15(B - V)^{3} +3.13(B - V)^{4})\times 10^{6} \ ,
\end{eqnarray}
\begin{eqnarray}\label{fit2R}
  \mathscr{F}_{\mathrm{IRT-2}}^{\mathrm{STELLA}} & = & (5.64 -13.32(B - V) + 17.15(B - V)^{2} \nonumber\\
  & & -11.57(B - V)^{3} +3.03(B - V)^{4})\times 10^{6} \ ,
\end{eqnarray}
\begin{eqnarray}\label{fit3R}
  \mathscr{F}_{\mathrm{IRT-3}}^{\mathrm{STELLA}} & = & (5.88 -13.61(B - V) + 16.60(B - V)^{2} \nonumber\\
  & & -10.58(B - V)^{3} +2.64(B - V)^{4})\times 10^{6} .
\end{eqnarray}

\subsection{Ultra-high-resolution PEPSI fluxes}

The IRT fluxes from the PEPSI spectra were again obtained the same way as for the RAVE and STELLA spectra. Giant targets were separately fit with the following fourth-order polynomials for the B--V colour term: 
\begin{eqnarray}\label{fit1P}
  \mathscr{F}_{\mathrm{IRT-1}}^{\mathrm{PEPSI}} & = & (5.65 -7.69(B - V) + 2.53(B - V)^{2} \nonumber\\
  & & +0.73(B - V)^{3} -0.39(B - V)^{4})\times 10^{6} \ ,
\end{eqnarray}
\begin{eqnarray}\label{fit2P}
  \mathscr{F}_{\mathrm{IRT-2}}^{\mathrm{PEPSI}} & = & (4.48 -7.66(B - V) + 4.78(B - V)^{2} \nonumber\\
  & & -1.20(B - V)^{3} +0.09(B - V)^{4})\times 10^{6} \ ,
\end{eqnarray}
\begin{eqnarray}\label{fit3P}
  \mathscr{F}_{\mathrm{IRT-3}}^{\mathrm{PEPSI}} & = & (5.11 -9.30(B - V) + 6.42(B - V)^{2} \nonumber\\
  & & -1.94(B - V)^{3} +0.21(B - V)^{4})\times 10^{6} .
\end{eqnarray}
These fits were obtained by excluding the following six targets: $\epsilon$~Eri, a well-known magnetically active star; 32~Gem, a bright A9 giant at 960\,pc with a variable $BV$ and spectral appearance; and four other targets (HD\,84937, HD\,103095, HD\,140283, HD\,122563) that are very metal-poor stars with [Fe/H] of between $-1.3$ and $-2.7$~dex.   

Observed 1-\AA\ IRT fluxes from the three instruments are compared for the one joint K3 target (HD\,82106) in Table~\ref{T:pepsi}. It quantifies what was already seen in Fig.~\ref{F-pepsicomp}b; that is, the higher the spectral resolution, the lower the 1-\AA\ flux. Fluxes obtained from STELLA and PEPSI for HD\,82106 were nearly equal and within its respective errors, whereas fluxes obtained from the RAVE spectrum were 11\% higher (for all its three IRT lines). We note that the ultra-high spectral resolution does not help significantly for the flux measurement (while it is essential for chromospheric studies based on the self-reversal of the emission-line core) but allows us to combine the high-resolution (dwarf) samples without further correction. 

\subsection{Combined STELLA plus PEPSI dwarf sample}

We now continue with the combined high-resolution sample, comprising the (continuum- and rotation-) corrected STELLA dwarfs and the PEPSI GBS dwarfs, and assume that it represents stellar radiative losses as close as possible to their non-magnetic expectation. This stellar sample is dubbed S+P. A polynomial fit of the fourth order is used to describe its average IRT fluxes versus B--V. The six GBS targets mentioned before of either very low metallicity, known magnetic activity, or that were a bright giant or supergiant were excluded from the S+P sample fit as well. The remaining targets have metallicities within approximately $\pm$0.5\,dex of the Sun, with a few outliers of mostly medium sub-solar metallicities (e.g.\ of $-0.75$ for HD\,6582 but with a large measurement uncertainty). The sample's rms scatter with respect to the polynomial fit is 16\%, and is larger at lower B--V and smaller at higher B--V. The fit parameters are:
\begin{eqnarray}\label{fit1P+S}
  \mathscr{F}_{\mathrm{IRT-1}}^{\mathrm{S+P}} & = & (8.36 -19.67(B - V) + 23.10(B - V)^{2} \nonumber\\
  & & -13.63(B - V)^{3} +3.12(B - V)^{4})\times 10^{6} \ ,
\end{eqnarray}
\begin{eqnarray}\label{fit2P+S}
  \mathscr{F}_{\mathrm{IRT-2}}^{\mathrm{S+P}} & = & (5.81 -13.95(B - V) + 17.50(B - V)^{2} \nonumber\\
  & & -11.16(B - V)^{3} +2.74(B - V)^{4})\times 10^{6} \ ,
\end{eqnarray}
\begin{eqnarray}\label{fit3P+S}
  \mathscr{F}_{\mathrm{IRT-3}}^{\mathrm{S+P}} & = & (6.02 -14.06(B - V) + 16.74(B - V)^{2} \nonumber\\
  & & -10.13(B - V)^{3} +2.38(B - V)^{4})\times 10^{6} .
\end{eqnarray}

\begin{figure}
  \centering
\includegraphics[width=\columnwidth]{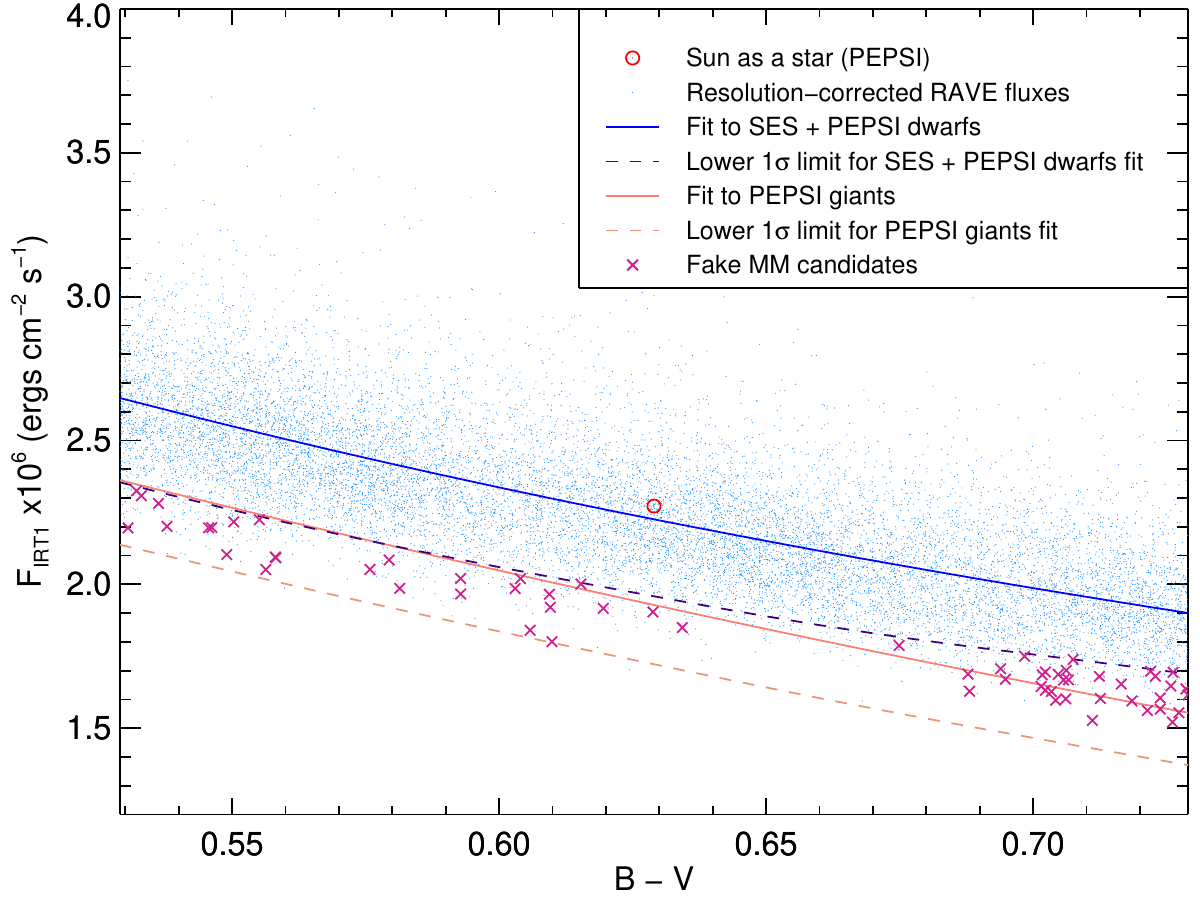}
  \caption{
  Resolution-corrected RAVE \ion{Ca}{ii} IRT-1 fluxes for a limited B--V range (14\,585 targets) and S/N$>$40. The continuous blue line is the fit to the high-resolution dwarf sample already shown in Fig.~\ref{Fallsamples}. The dashed blue line is its 1$\sigma$ lower bound. Also shown is the fit to the GBS-giants from PEPSI spectra (continuous orange line) and its 1$\sigma$ lower bound for comparison. The targets within the two dashed lines were additionally constrained by requiring them to fall below the upper bound in all three IRT lines and are highlighted as crosses (and listed in the Appendix in Table~\ref{TappMMcands}). All except one of these candidates turned out to exhibit supersolar metallicity. We thus called them fake MM candidates.
  }
  \label{F5}
\end{figure}

The {\it Gaia} DR3 radiative losses (photospheric corrected; $\log R'_{\rm IRT}$) presented by 
\citet{2023A&A...674A..30L} 
range from a low of $-8.0$ to a high of $-4.5$~W\,m$^{-2}$ (they use the SI system), with the bulk of stars at $-5.7$~W\,m$^{-2}$ around $T_{\rm eff} \approx 5800$\,K. It appears that 
\citet{2023A&A...674A..30L}
truncated their sample intentionally at a lower limit of $\log R'_{\rm IRT}=-8.0$ W\,m$^{-2}$ (--5.0\,\ecms\ in the cgs system). This is to be compared to our PEPSI solar spectrum once corrected for a photospheric contribution. Using the corrections provided by 
\citet{2017A&A...605A.113M}, 
our fluxes convert to a solar IRT radiative loss of $\log R'_{\rm IRT} = -5.00$\,\ecms\ (or $-8.00$ W\,m$^{-2}$), which accidentally exactly equals the truncation limit of the {\it Gaia} DR3 data release. The PEPSI solar spectrum was taken in December 2016, approaching the solar activity minimum in 2018--2020. The monthly mean Sunspot number in December 2016 was then 21.4 (a minimum of 0.5 in December 2019, and a maximum of 150 in February 2014). This PEPSI measurement thus represents a comparable low activity state of the Sun following the ISES solar-cycle progression NOAA website\footnote{https://www.swpc.noaa.gov/}. Such (very) low IRT fluxes were excluded in the sample of 
\citet{2023A&A...674A..30L}.
This basically means that we can only partially compare our minimum-flux stars or MM-candidate stars to the targets in the {\it Gaia} DR3 data release. 

\section{Analysis and discussion}\label{analysis}

\subsection{Towards a MM-candidate sample}\label{S51}

Our first analysis assumption is now that our combined high-resolution S+P dwarf-star sample represents low-activity stars with otherwise evolutionarily normal IRT emission-line fluxes. We thereby implicitly assume that fully inactive (cool) stars do not exist. On the other hand, the large number of available RAVE targets makes it most likely that its average (or median) also represents such low-activity stars with normal IRT fluxes, which is our second working assumption. With these two assumptions, systematic differences in IRT flux between the RAVE and S+P samples at a given B--V must mostly be due to the different spectral resolution, plus the unavoidable individual stellar intrinsic differences (the latter mostly due to metallicity and the snapshot observations in the presence of undetected rotational modulation and cycles). The latter assumption is supported by the fact that when high-resolution spectra are degraded to the RAVE resolution, the measured fluxes remain the same.

By shifting the average RAVE fluxes (e.g.\ for IRT-1 represented by Eq.~\ref{fit1}) to the respective average high-resolution STELLA+PEPSI fluxes (e.g.\ for IRT-1 represented by Eq.~\ref{fit1P+S}), we removed the expected resolution effect from the RAVE fluxes to the first order. These fluxes were then used in the further analysis in this paper. Furthermore, considering the respective sample scatter in an additive way,
\begin{equation}
    \sigma = \sqrt{\sigma_{\mathrm RAVE}^2 + \sigma_{\mathrm S+P}^2}
,\end{equation}
we defined a lower bound for a (resolution-corrected) RAVE observation of a `normal' low-activity dwarf flux:  
\begin{equation}\label{eq_corrflux}
    \mathscr{F}_{\mathrm{corr}}^{\mathrm{RAVE}}  =  \mathscr{F}_{\mathrm{obs}}^{\mathrm{RAVE}} - (\mathscr{F}_{\mathrm{fit}}^{\mathrm{RAVE}} - \mathscr{F}_{\mathrm{fit}}^{\mathrm{S+P}}) - n \ \sigma \ .
\end{equation}
This lower bound is shown as a dashed (blue) line in Fig.~\ref{F5} for a $n=1$ assumption (and again in the Appendix for the other IRT lines). Any RAVE observation with a corrected IRT flux lower than that may be defined as a MM-candidate star. 

We limited the initial RAVE dwarf sample to a B--V range $\pm$0.1 around the solar value, that is B--V between 0.53--0.73, or $T_{\rm eff}$ between approximately 6200--5400\,K (F8--G8). With these two criteria, the flux limit and B--V limit, we identify a total of 598, 244, and 358 targets below our lower bound from IRT-1, IRT-2, and IRT-3, respectively. The further assumption, that only targets are considered for which the fluxes from all three IRT lines fall below the flux limit, reduces the three samples to a joint 61 targets. The spectra of these targets were then visually inspected towards leftover observational issues (see Sect.~\ref{reject} in the Appendix); for example, a combination of bad S/N, continuum displacement, wavelength shift, and others (a.o.). For this, we adopted a more stringent S/N requirement than in the initial rejection by considering targets only if S/N$>$40. We note that if applied to all RAVE spectra it downsizes the overall sample from 78\,111 to 47\,618 dwarf stars. However, none of the 61 targets were affected by any of the obvious instrumental defects and we highlight these targets in Fig.~\ref{F5} as crosses. 

This sample of 61 candidate stars appears with $\log g$ errors of $\pm$0.12--0.16, and [Fe/H] errors of $\pm$0.10--0.12. None of these targets have `exceptional' large errors that would question their inclusion. At this point, we recall that we had applied a gravity-based threshold for the initial dwarf sample of $\log g$ being larger than 4.0. The actual range for the 61-target sample turned out to be between 4.01--4.65. However, all but one target appear with a supersolar metallicity in the range +0.10 to +0.50, with a sample average of +0.35, well above the RAVE ensemble uncertainty of at most 0.16\,dex. Only one target (J134102.1-214222) has [Fe/H] of --0.20. Obviously, the high metallicity faked the (relative) flux measurement. Therefore, we dub this 61-target sample our fake MM sample. Table~\ref{TappMMcands} in the Appendix lists these 61 targets along with their stellar parameters from RAVE DR6. In the next paragraph, we attempt to correct the IRT fluxes for metallicity before extracting our final MM candidates. 

\subsection{Correction for metallicity}

Higher metallicity makes the IRT absorption lines deeper, and thus the flux in the line cores apparently lower, although the depression is generally smaller in the line core than in  the wings and smaller for cooler stars than for hotter stars 
\citep[see, e.g.][]{2000A&A...353..666C}. 
Lower metallicity, on the contrary, makes the Ca lines shallower and the line-core fluxes higher. As for the rotational-broadening correction in Sect.~\ref{broad}, we again followed an empirical approach for its correction based on the measured metallicity from the RAVE data-analysis pipeline in DR6. Individual error bars are non-existent and some caution exists, as was already noted in DR6 by
\citet{2020AJ....160...83S, 2020AJ....160...82S}. 
Nevertheless, the metallicity ensemble errors are well constrained. Our RAVE dwarf sub-sample with S/N$>$40 and B--V=0.53--0.73 has an average (ensemble) metallicity error of 0.11\,dex. \citet{2000A&A...353..666C} noted that even the noise amplitude for a S/N=100 spectrum corresponds for $T_{\rm eff}$=5500\,K to a change in metallicity of 0.15\,dex. 

\begin{figure}
{\bf a.}\\
  \includegraphics[angle=0,width=8.4cm, clip]{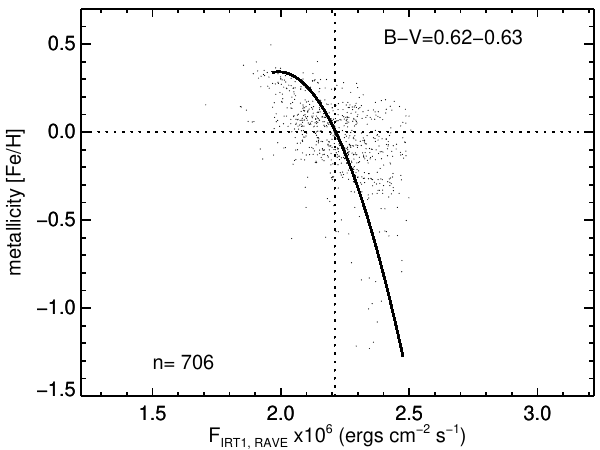}

{\bf b.}\\
\includegraphics[angle=0,width=8.4cm, clip]{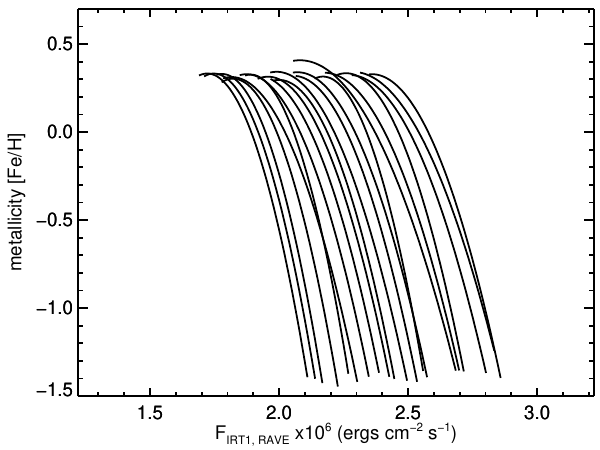}

{\bf c.}\\
  \includegraphics[angle=0,width=8.4cm, clip]{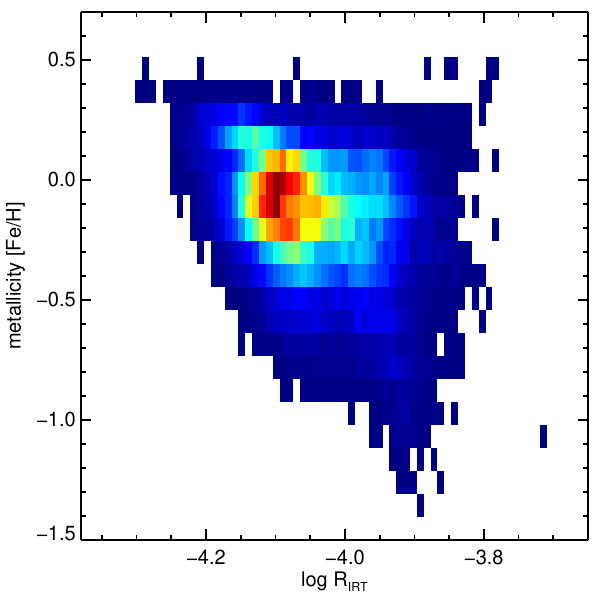}
   \caption{
   Dependency of Ca\,{\sc ii} IRT flux on metallicity. $a.$ Example relation for a colour bin of B--V=0.62--0.63 for IRT-1 (log-linear presentation). Small dots are the data of 706 targets in this B--V bin. The line is a polynomial fit of the second order. $b.$ Overplot of the polynomial fits for all B--V bins. $c.$ Logarithmic radiative loss from the entire triplet and for the full available B--V range. The colour denotes the number density (darkest red is 600, darkest blue is 1 target). 
   }
   \label{F6}
\end{figure}

For our correction procedure, we first separated the B--V=0.53--0.73 sub-sample into narrow colour bins of 0.01~mag  ($T_{\rm eff}\approx 50-100$\,K) and then fitted a simple linear regression (SLR) to $\mathscr{F}_{\mathrm{IRT}}$ as $f$([Fe/H]) in each colour bin. In order not to bias these fits with stars of metallicity-unrelated higher fluxes, that is the chromospherically more active stars, we excluded targets from the SLR fit that have fluxes higher than 1$\sigma$ of the bin-sampled average. Because we did not know metallicity errors for individual targets, we did not use the SLR fits directly for the metallicity corrections in order not to over correct the data. Instead, we proceeded with a simple bin-averaging strategy. For this, we first defined in each B--V bin the average flux for targets with measured [Fe/H]=0 (actually a bin within $\pm$0.05\,dex). Secondly, we repeated this for the range $+0.5\leq$[Fe/H]$\leq -0.5$ (again always binned into groups of 0.1\,dex), and computed their average fluxes for each bin. Thirdly, the difference between these fluxes and the flux for [Fe/H]=0 is defined as our correction for metallicity within a given B--V bin: 
\begin{equation}\label{Mcorr}
    \delta \mathscr{F}_{\mathrm{obs}}^{\mathrm{corr}} = \mathscr{F}_{\mathrm{Fe/H=0}} - \mathscr{F}_{\mathrm{Fe/H}}
,\end{equation}
with $\mathscr{F}_{\mathrm{Fe/H}}$ being the flux from a  polynomial fit to the individual data,
\begin{equation}
[\mathrm{Fe/H}] = a + b \mathscr{F}_{\mathrm{Fe/H}} + c \mathscr{F}_{\mathrm{Fe/H}}^2 \ .
\end{equation}
The zero-metallicity flux, $\mathscr{F}_{\mathrm{Fe/H=0}}$, and the polynomial-fit coefficients, $a$, $b$, and $c$, are collected as a function of B--V in the Appendix in Table~\ref{TAppMetal}. The corrections from Eq.~(\ref{Mcorr}) can be positive or negative and are of an amount up to 14\%\ of the nominal zero-metallicity flux, but typically only half of that. We note that the comparably few targets with metallicities of less than --0.5\,dex were combined for the determination of the metallicity correction (then a bin of 1.0\,dex) because otherwise the binning would have been severely under-sampled. There were no targets with [Fe/H]$>$+0.50 in our B--V restricted sample.  

Figure~\ref{F6}, panel $a$, shows an example  (for $\mathscr{F}_{\mathrm{IRT-1}}$) for the colour bin including the Sun-as-a-star, B--V=0.62--0.63, and its second-order polynomial fit. Its total number of stars, 722, is reduced due to the above +1$\sigma$ exclusion rule to 706. It is obvious that lower-metallicity targets exhibit systematically higher relative fluxes than solar- and supersolar metallicity stars, as was expected. Panel~$b$ shows the polynomial fits from all B--V bins. We note that its spread just reflects the B--V dependency of the photospheric contribution. Figure~\ref{F6}, panel~$c$, plots the entire sample of 47\,618 stars with B--V ranging from 0.3 to 1.5 versus the radiative loss $\log R_{\rm IRT}$; that is, the sum of fluxes scaled with the bolometric luminosity instead of $\mathscr{F}_{\mathrm{IRT-1}}$. We see a similar but blurred metallicity tendency as for the individual line fluxes.

\begin{figure}
  \centering
\includegraphics[width=\columnwidth]{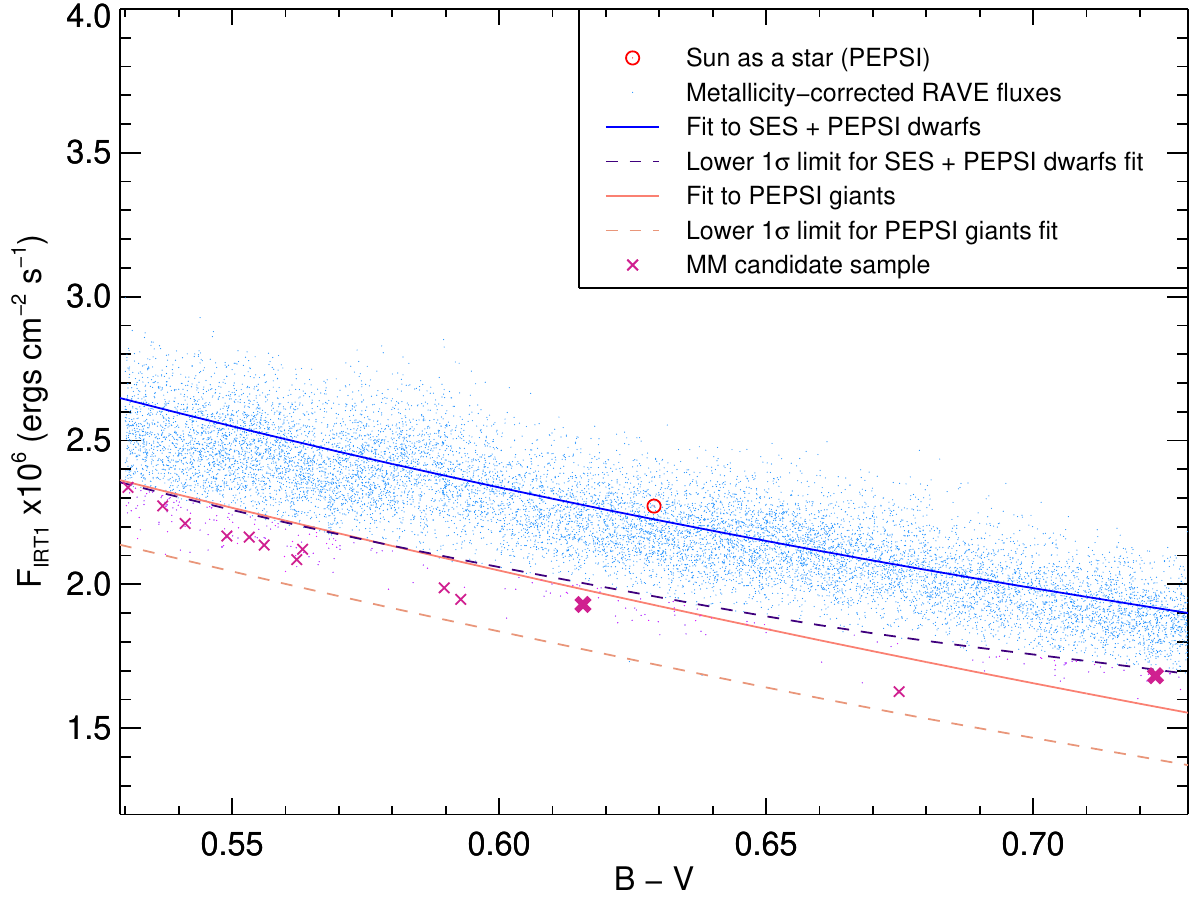}
  \caption{
  Metallicity-corrected RAVE \ion{Ca}{ii} IRT-1 fluxes (13\,326 targets). Otherwise, as in Fig.~\ref{F5} (see Appendix for the other two IRT lines). Crosses are the possible candidate targets from Table~\ref{FinalMMcands}. Boldfaced crosses are the two MM candidates, TIC\,352227373 (RAVE J191213.1-760732) and TYC~7560-477-1 (RAVE J025410.6-383603).
  }
  \label{F7}
\end{figure}

Finally, Fig.~\ref{F7} shows the result of repeating the same procedure described in the previous section \ref{S51} but for the metallicity-corrected sample (13\,326 stars). Targets below the original $1\sigma$ lower bound (dashed blue line in Fig.~\ref{F7}) in all three IRT lines now compromise our final MM candidates. These MM candidates show IRT fluxes between 13--22\% below the solar minimum. When looking at different IRT lines individually, the number of candidates varies: with IRT-1 line, we find 250; with IRT-2, 84; and with IRT-3, 124. Further, the IRT-1 and IRT-2 lines have 29 common candidates, IRT-1 and IRT-3 have 22, and IRT-2 and IRT-3 have 24. However, only 13 targets are identified as candidates when using all three IRT lines. Four targets remained from the fake sample.

\begin{table*}
\caption{Final sample of lowest-activity targets in the B--V range 0.53--0.73 from the RAVE survey. }\label{FinalMMcands}
\begin{small}
\begin{tabular}{lccrclllllll}
\hline\hline \noalign{\smallskip}
RAVE name & $T_{\rm eff}$ & $\log g$  & [Fe/H] & (B--V)$_0$ & $\mathscr{F}_{\mathrm{IRT-1}}$ & $\mathscr{F}_{\mathrm{IRT-2}}$ & $\mathscr{F}_{\mathrm{IRT-3}}$ & $\log R_{\rm IRT}$ & $L$ & $R$ & Other \\
     & (K)          & (dex)     & (dex)  & (mag)  & \multicolumn{3}{c}{($10^6$ \ecms)} & ($\sigma T_{\rm eff}$) & (L$_\odot$) & (R$_\odot$) & name \\
\noalign{\smallskip}\hline \noalign{\smallskip}
J091344.0-185715 & 6078 & 4.07 & $-$0.06 & 0.53 & 2.3369 & 1.5231 & 1.5991 & $-$4.15 & 8.2 & 2.6 & TYC 6036-1085-1 \\
J210322.6-374731 & 6115 & 4.02 & $+$0.11 & 0.54 & 2.2724 & 1.5224 & 1.6340 & $-$4.16 & 5.5 & 2.1 & CPD-38 8173$^a$ \\
J210118.8-604737 & 6027 & 4.19 & $-$0.11 & 0.54 & 2.2114 & 1.5028 & 1.5405 & $-$4.15 & 4.1 & 1.9 & CPD-61 6520 \\
J200531.9-522958 & 6243 & 4.62 & $+$0.42 & 0.55 & 2.1681 & 1.3992 & 1.5136 & $-$4.23 & 2.6 & 1.4 & TYC 8404-7-1$^b$ \\
J193243.7-551333 & 5971 & 4.17 & $-$0.15 & 0.55 & 2.1644 & 1.4197 & 1.5814 & $-$4.14 & 5.5 & 2.2 & TIC 320005130 \\
J014458.2-472433 & 5837 & 4.02 & $-$0.52 & 0.56 & 2.1365 & 1.4453 & 1.5049 & $-$4.11 & 3.3 & 1.8 & TYC 8040-22-1 \\
J024718.6-811225 & 5936 & 4.09 & $-$0.14 & 0.56 & 2.0863 & 1.3579 & 1.5066 & $-$4.15 & 6.7 & 2.5 & TYC 9374-1033-1 \\
J212620.6-031522 & 6010 & 4.05 & $+$0.08 & 0.56 & 2.1221 & 1.4442 & 1.5375 & $-$4.16 & 7.4 & 2.5 & TYC 5199-577-1 \\
J225620.6-473754 & 5680 & 4.09 & $-$0.68 & 0.59 & 1.9878 & 1.3423 & 1.3637 & $-$4.10 & 43 & 6.8 & CPD-48 10837 \\
J134102.1-214222 & 5816 & 4.15 & $-$0.20 & 0.59 & 1.9479 & 1.2159 & 1.4459 & $-$4.15 & 8.0 & 2.8 & CPD-21 5592$^{a,b}$ \\
J191213.1-760732 & 5786 & 4.72 & $-$0.11 & 0.62 & 1.9303 & 1.3100 & 1.2598 & $-$4.15 & 1.16 & 1.07 & TIC 352227373 \\
J055244.5-404031 & 5812 & 4.33 & $+$0.50 & 0.67 & 1.6271 & 1.0293 & 1.1135 & $-$4.23 & 2.3 & 1.5 & TYC 7602-148-1$^b$ \\
J025410.6-383603 & 5509 & 4.18 & $+$0.10 & 0.72 & 1.6822 & 1.0943 & 1.0672 & $-$4.13 & 0.98 & 1.1 & TYC 7560-477-1$^b$ \\
\hline
\end{tabular}
\end{small}
\tablefoot{$^a$Double- or Multiple star. $^b$Also in the sample of fake candidates in Table~\ref{TappMMcands}. $T_{\rm eff}$, $\log g$, and [Fe/H] are from RAVE DR6. (B--V)$_0$ was converted from $T_{\rm eff}$. The next four columns are the resolution- and metallicity-corrected fluxes for each of the IRT lines and the combined radiative loss in logarithmic form, respectively. The luminosity, $L$, and radius, $R$, are from the DR3 parallax and the Simbad visual magnitude corrected for interstellar absorption according to
\citet{2011ApJ...737..103S}. 
The distance to CPD-21$^\circ$5592 was assumed to be 350\,pc. }
\end{table*}

\subsection{The Sun in context}

\citet{2022AN....34323996D}
characterised the \ion{Ca}{ii} H\&K activity of the Sun with PEPSI spectra as in this paper. They found an average S-index of 0.1522$\pm$0.0003 for 2018--19, with values varying between 0.152 and 0.153 due to rotational modulation. This average was obtained during the declining phase of the weak solar cycle~24 and is about 7\% lower than the S-value of 0.163 reported by 
\citet{2017ApJ...835...25E}
for the minimum between solar cycles 23 and 24 in December 2008. Slightly lower S-values were measured for the  PEPSI spectra from November 16 and 17 2016 and resulted in S-index values of 0.148 and 0.150, respectively (see fig.~4 in
\citet{2022AN....34323996D} 
along with measurements from various solar atlases). Comparable low values were measured from Moon spectra by 
\citet{2012A&A...540A.130S},
0.150 on plage-free days, but slightly larger values were also reported, for example, by 
\citet{2019A&A...627A.118M}
based on HARPS spectra that were 4.6\% higher than the respective average PEPSI S-index. The Sun has been observed using the IRT-2 line for several decades 
\citep{2010MmSAI..81..643L,2024FrASS..1028364Z}. 
In particular, 
\citet{2024FrASS..1028364Z}
showed that the correlation of different indices, including the IRT-2 line, with solar magnetic activity changes with time, being stronger during maximum solar activity. Therefore, we must expect not only cycle-dependent S-index variations of a few percent but also B--V-dependent additional (basal) activity. 

Based on these ultra-high-resolution `Sun-as-a-star' spectra from November 2016, we can directly relate the solar H\&K S-index from 
\citet{2022AN....34323996D} 
with our (solar) IRT fluxes: the average of $S=0.149$ relates to an IRT-1 flux of 2.305$\times$10$^6$\,\ecms\ (IRT-2: 1.636$\times$10$^6$, IRT-3: 1.686$\times$10$^6$) or $\log R_{\rm IRT}=-4.05$, the latter photospherically uncorrected. This is plotted with the solar symbol in Fig.~\ref{F5} and shows that the Sun falls within the bulk of the RAVE stars once their fluxes are corrected for spectral resolution. Approximately half of the RAVE targets in the solar B--V bin are below this minimum solar flux. At this point, we remark that it is not the intention of this paper to define a globally valid H\&K versus IRT relationship because it would require truly accurate estimates of the photospheric contribution, which is left for future work (instead, we refer to the work of 
\citet{2018A&A...616A.108B,2023A&A...674A..30L},
a.o.). However, our working hypothesis is that a MM is an activity-free phase, while even plage-free days on today's Sun are not strictly activity free.  After all, the Sun is currently not in a MM. Therefore, adopting the all-time lowest measured S-index of the Sun for its MM-value is likely an overestimation. At this point, we emphasise that the concept of basal chromospheric flux 
\citep{1995A&ARv...6..181S} 
still lacks a detailed physical implementation, even for the Sun 
\citep{2012A&A...540A.130S}, 
which then could provide a theoretical value. We are thus more confident that our conservative approach in Eq.~\ref{eq_corrflux} is currently better suited for a selection limit. It sets a (photospherically uncorrected) bona fide MM IRT-1 flux for solar $T_{\rm eff}$ to $1.95\times$10$^6$\,\ecms\ (IRT-2: 1.32$\times$10$^6$, IRT-3: 1.38$\times$10$^6$) or $\log R_{\rm IRT}=-4.13$ (the latter $\approx$20\% below the November 2016 PEPSI solar minimum flux). 

\subsection{\ion{Ca}{ii} IRT occurrence among late-type stars}

Because the RAVE-survey spectra are from one-time visits, we cannot compare them with the cycle morphology that came out of, for example, the MWO survey. However, we can compare the occurrence rate in flux and colour bins with those in the MWO and the Lowell-HAO SSS (solar-stellar spectrograph) data.
\citet{1990Natur.348..520B}
found a bimodal distribution of the \ion{Ca}{ii} $S$-indices in the MWO sample -- with higher-activity and lower-activity components -- and described this with two Gaussian envelopes centred at $S$=0.145 and 0.170, respectively. They suggested that the lower-activity component is compromised of stars sampled in a MM-like phase, and defined this as \ion{Ca}{ii} $S$-values below the solar minimum (0.164) and with a flat time series ($\sigma\leq$ 1.5\%). The lowest observed value for a star with $T_{\rm eff}\approx$solar in the MWO survey is 0.140. This may constitute the non-magnetic contribution to the radiative losses in \ion{Ca}{ii}~H\&K
(i.e.\ Schrijver's
\citeyearpar{1995A&ARv...6..181S}
`basal flux').
\citet{1990Natur.348..520B}
then suggested that cycling stars can undergo a transition to a MM phase with activity as little as 25\%\ of the Sun during solar minimum. Although
\citet{1992ASPC...27..150S}
revised the original estimate of the number of MM stars from 30\%\ to 10--15\%, Baliunas and Jastrow's suggestion had a far-reaching impact and spurred a number of studies aimed at understanding the acoustic and magnetic heating of such stars
\citep{1994ApJ...427L.111Z, 1998ASPC..154..211S, 2004ApJ...609..392J}.
Later, however,
\citet{2004ApJ...614..942H}
could not verify the bimodality from independent Lowell-SSS data and noted that a flat activity state does not necessarily imply weak (i.e.\ subsolar) magnetic activity. 

\begin{figure}
{\bf a.}\\
   \includegraphics[width=8.5cm]{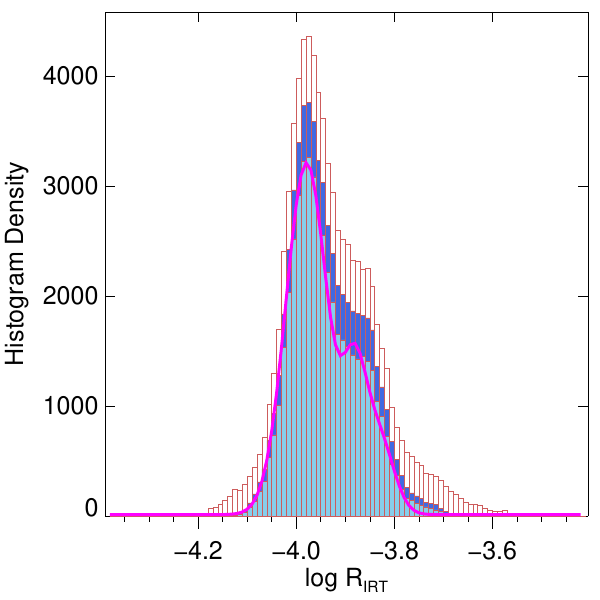}
   
{\bf b.}\\
   \includegraphics[width=8.5cm]{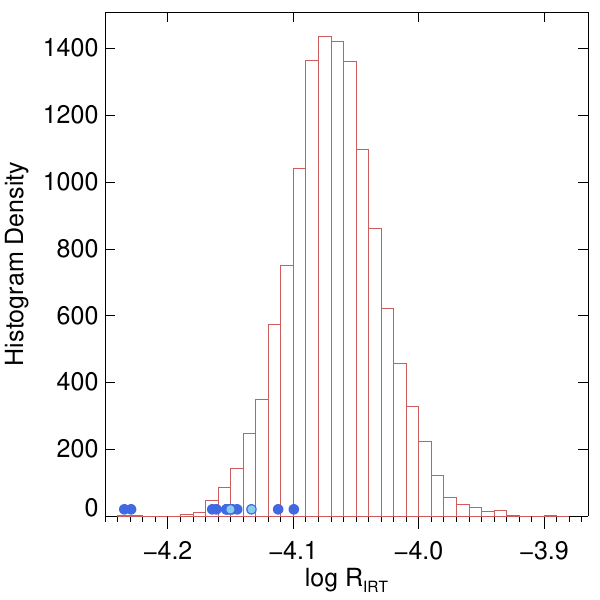}
   \caption{Number density versus (photospheric uncorrected) radiative loss, $\log R_{\rm IRT}$, from RAVE data. $a.$ Density distribution from the resolution-corrected but metallicity-uncorrected sample. The three colours indicate increasing average S/N. White is for the initial RAVE targets with spectra of S/N$>$15, dark blue is for a sample with S/N$>$30, and light blue for S/N$>$40. The magenta line is a double Gaussian fit to the sample with S/N$>$40. $b.$ As above, but with metallicity correction and for B--V limited to 0.53--0.73~mag. The radiative losses of the 13 stars from Table~\ref{FinalMMcands} are marked with blue dots. The two MM candidates are highlighted with light blue.}
   \label{F8}
\end{figure}

Figure~\ref{F8}a shows the number distribution of  radiative losses for the resolution-corrected but metallicity-uncorrected RAVE dwarf-star sample (full available B--V range). Three sample variants are plotted to demonstrate the impact of S/N on the morphology of the low-activity tail, and whether we also see a bimodality. The initial sample had been made up from spectra with S/N$>$15 (78,111 targets). The other two samples are with spectra of S/N$>$30 (58\,672 targets) and S/N$>$40 (47\,618 targets). Figure~\ref{Rn} shows the number distribution of the resolution- and metallicity-uncorrected fluxes for the same three samples for comparison. We emphasise that the low-activity peak seen in the flux distribution in Fig.~\ref{Rn} at $\approx$1$\times$10$^6$\,\ecms\ in the initial S/N$>$15 sample basically disappeared in the higher S/N samples and is thus an artefact. No such peak exists in any of the three radiative-loss distributions in Fig.~\ref{F8}a. However, a clear secondary peak is identified at higher radiative losses. A two-Gaussian fit to the S/N$>$40 sample identifies the main number maximum at $\log R_{\rm IRT}= -3.97$ and a secondary peak at $\log R_{\rm IRT}= -3.88$. The latter comprises the sum of the hotter stars with comparably higher photospheric fluxes, and the cooler stars with high chromospheric activity.  

Figure~\ref{F8}b shows the number distribution of our solar-like stars from the S/N$>$40 sample with B--V in the range 0.53--0.73. Here, we only plot the radiative losses from the metallicity-corrected fluxes; that is, the best fluxes that we could extract from the RAVE data. Surprisingly, its distribution does not differ notably  from a Gaussian, indicating the lack of a very-low-activity MM-star peak as interpreted from the MWO data by  \citet{1990Natur.348..520B}. 

\subsection{The final trap to MM candidacy: Off the main sequence}

The angular-momentum evolution generally suggests lower activity levels for evolved stars than for main-sequence stars. An unusually low \ion{Ca}{ii} $S$ index or IRT flux of a (mildly) evolved star without a known luminosity could therefore be misinterpreted as being due to the MM state of a main-sequence star.
\citet{2004AJ....128.1273W}
found that most of the stars previously identified as MM stars were in fact older stars that had already evolved off the main sequence. Wright's sample was then based on the $S$-index measurements from the California and Carnegie Planet Search program
\citep{2004ApJS..152..261W},
the MWO program, the
\citet{1996AJ....111..439H}
`Phoenix' survey (SETI successor), and the MM candidates from
\citet{2003AJ....126.2048G},
all of them combined with the published parallaxes from \emph{Hipparcos}
\citep{1997ESASP1200.....E}. 

Table~\ref{FinalMMcands} lists the 13 RAVE stars that we found with the lowest IRT fluxes in the B--V range 0.53--0.73 after correcting for metallicity. Figure~\ref{Fi:MM_13_sample} plots their spectra. The Gaia DR3 distances for 11 of them suggest that they must be evolved stars. With the visual magnitudes taken from Simbad, and the proper interstellar absorption taken from \citet{2011ApJ...737..103S}, we obtain luminosities for these 11 stars of between 43 and 2.3 times solar. Therefore, we dismiss 11 of the 13 targets as MM candidates. The remaining two stars appear with luminosities of 1.16~L$_\odot$ and 0.98~L$_\odot$ (with $T_{\rm eff}$ of 5786\,K and 5509\,K, respectively) and are thus likely main-sequence dwarfs. Its respective (luminosity) uncertainties are driven by the errors of the effective temperature plus the amount of interstellar-absorption correction in general, and likely amount to $\approx$10--20\%.    


\section{Conclusions}\label{sec:conc}

Subsolar and supersolar metallicities fake relative chromospheric fluxes that are too large or too low, respectively, if measured the standard way as an equivalent width. Metallicity is thus an important parameter for the final line-flux determination, in particular for very low activity levels. It also impacts indirectly on the age-activity relation itself, and must not be neglected anymore. 

The small number of very-low-activity stars from our metallicity-corrected \ion{Ca}{ii} IRT flux sample in Fig.~\ref{F7}, 13 candidates out of the 13\,326 main-sequence F8--G8 targets, indicates that stars in a Maunder-like minimum must be very rare. This means that at most $\approx$0.1\% of the main-sequence stars in the B--V range of $\pm$0.1 of the Sun show levels of magnetic activity significantly ($\approx$20\%) below solar minimum. Even more devastatingly, 11 of these 13 stars have Gaia DR3 parallaxes and visual brightnesses that identify them as evolved stars rather than main-sequence dwarfs. We conclude that the remaining two dwarf stars, TIC~352227373 (G2V) and TYC~7560-477-1 (G7V), are the only two MM candidate stars from the present search.  


\begin{acknowledgements}
We thank the anonymous referee for their comments.\\
S.P.J. acknowledges support from
\emph{Deut\-sche For\-schungs\-ge\-mein\-schaft, DFG} project number JA~2499/1--1.\\
Funding for RAVE has been provided by: the Australian Astronomical Observatory; the Leibniz-Institut fuer Astrophysik Potsdam (AIP); the Australian National University; the Australian Research Council; the French National Research Agency; the German Research Foundation (SPP 1177 and SFB 881); the European Research Council (ERC-StG 240271 Galactica); the Istituto Nazionale di Astrofisica at Padova; The Johns Hopkins University; the National Science Foundation of the USA (AST-0908326); the W. M. Keck foundation; the Macquarie University; the Netherlands Research School for Astronomy; the Natural Sciences and Engineering Research Council of Canada; the Slovenian Research Agency; the Swiss National Science Foundation; the Science \& Technology Facilities Council of the UK; Opticon; Strasbourg Observatory; and the Universities of Groningen, Heidelberg and Sydney. The RAVE web site is at https://www.rave-survey.org. 
Based partially on data obtained with the STELLA robotic telescopes in Tenerife, an AIP facility jointly operated by AIP and IAC and funded by the Brandenburg Ministry of Science and Culture. The STELLA web page is at https://stella.aip.de/ . 
Based partially on data acquired with the Large Binocular Telescope (LBT). The LBT is an international collaboration among institutions in the United States, Italy, and Germany. LBT Corporation partners are the University of Arizona on behalf of the Arizona university system; Istituto Nazionale di Astrofisica, Italy; LBT Beteiligungsgesellschaft, Germany, representing the Max-Planck Society, the Leibniz-Institute for Astrophysics Potsdam (AIP), and Heidelberg University; the Ohio State University; and the Research Corporation, on behalf of the University of Notre Dame, University of Minnesota and University of Virginia. The PEPSI web site is at https://pepsi.aip.de/ .\\
This research has made use of the SIMBAD database, operated at CDS, Strasbourg, France, which is greatly appreciated.\\
\end{acknowledgements}


  \bibliographystyle{aa} 
   \bibliography{rave} 


\begin{appendix}

\section{Initial rejection criteria for RAVE spectra}\label{reject}

\begin{figure}
\centering
\includegraphics[width=.8\columnwidth]{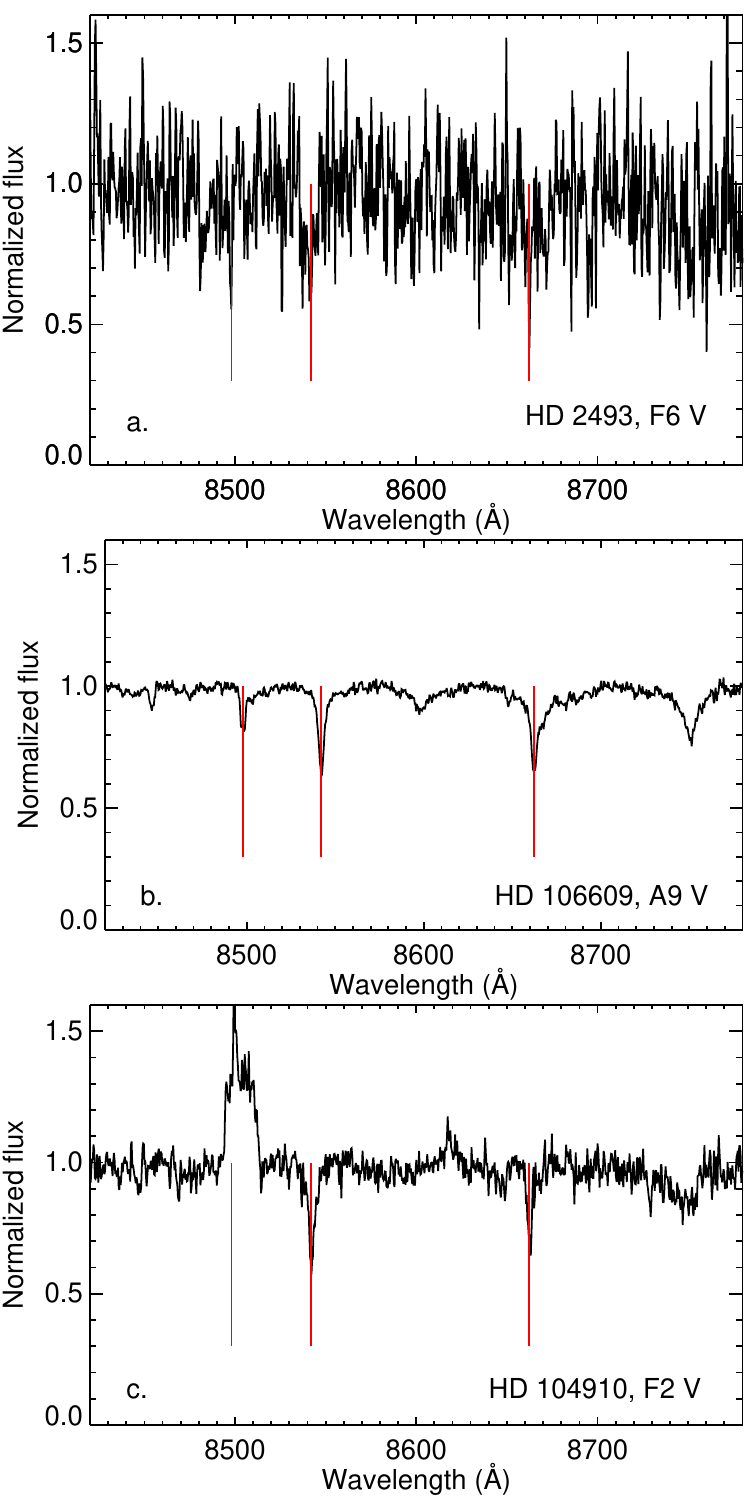}
\caption{
Three examples illustrating the reasons why some RAVE spectra were rejected:
  a) A spectrum of the F6 dwarf
  \citep{1975mcts.book.....H}
  HD\,2493 illustrating a case of low S/N.
 b) A spectrum of the A9 dwarf
 \citep{1988mcts.book.....H}
 HD\,106609 illustrating contamination by Paschen lines.
 c) A spectrum of the F2 dwarf
 \citep{1988mcts.book.....H}
 HD\,104910 illustrating a case of odd feature seen in the spectrum.
\ion{Ca}{ii} IRT lines are marked with red vertical lines.
          }
      \label{AFi:bad}
\end{figure}

Low S/N (S/N$<$15 per pixel) was a reason to reject 6,648 ($7.6\%$) spectra. In many cases it also led to not well defined continuum levels, and therefore to a wrong equivalent width measurement. An example of a rejected spectrum based on bad S/N is shown in Fig.~\ref{AFi:bad}a.

Blending with strong and broad Paschen lines was a reason to reject 1,094 ($<1.3\%$) spectra. An example of a rejected spectrum is shown in Fig.~\ref{AFi:bad}b.

In addition to that, 862 spectra (<1\%) were rejected due to various reasons. Those were odd features in the spectra, problems with continuum level or wavelength calibration. Also in same cases binarity or multiplicity was evident. An example showing an odd looking spectrum is shown in Fig.~\ref{AFi:bad}c.

\FloatBarrier

\section{Additional data analysis and figures}

\subsection{Definition of radiative loss in the IRT lines}

In order to catch all of the radiative loss in the RAVE IRT lines, we base our comparisons on the sum of the corrected  fluxes from all three IRT lines and express them in units of the bolometric luminosity:
\begin{equation}\label{eqRIRT}
R_{\rm IRT} =
\frac{\mathscr{F}_{\rm IRT-1} + \mathscr{F}_{\rm IRT-2} + \mathscr{F}_{\rm IRT-3}}{\sigma \ T_{\rm eff}^4} , 
\end{equation}
in analogue to the \ion{Ca}{ii} H\&K radiative loss introduced by
\citet{1979ApJS...41...47L}:
\begin{equation}\label{eqRCaHK}
R_{\rm HK} =
\frac{\mathscr{F}_{\rm H} + \mathscr{F}_{\rm K}}{\sigma \ T_{\rm eff}^4} .
\end{equation}
We recall that the IRT fluxes and radiative losses in the present paper are not corrected for photospheric contribution (otherwise indicated by a prime for the respective parameter) while generally the H\&K based values in the literature are. However, for low-activity stars such corrections are on the same order of the measured total fluxes. 
\citet{2017A&A...605A.113M}
provided a convenient collection of such corrections based on 26 bona fide inactive stars as a function of B--V and $v\sin i$. Its numerical corrections are between 4$\times$10$^6$ to 7$\times$10$^6$\,\ecms\ for the sum of all three IRT lines for B--V of 0.8 to 0.5, respectively. Regarding a discussion of the use of a photospheric correction, we refer to the Appendix in the work of 
\citet{2023A&A...674A..30L}. 

\subsection{Supplementary figures}

\begin{figure*}
\centering
  \includegraphics[width=.43\textwidth]{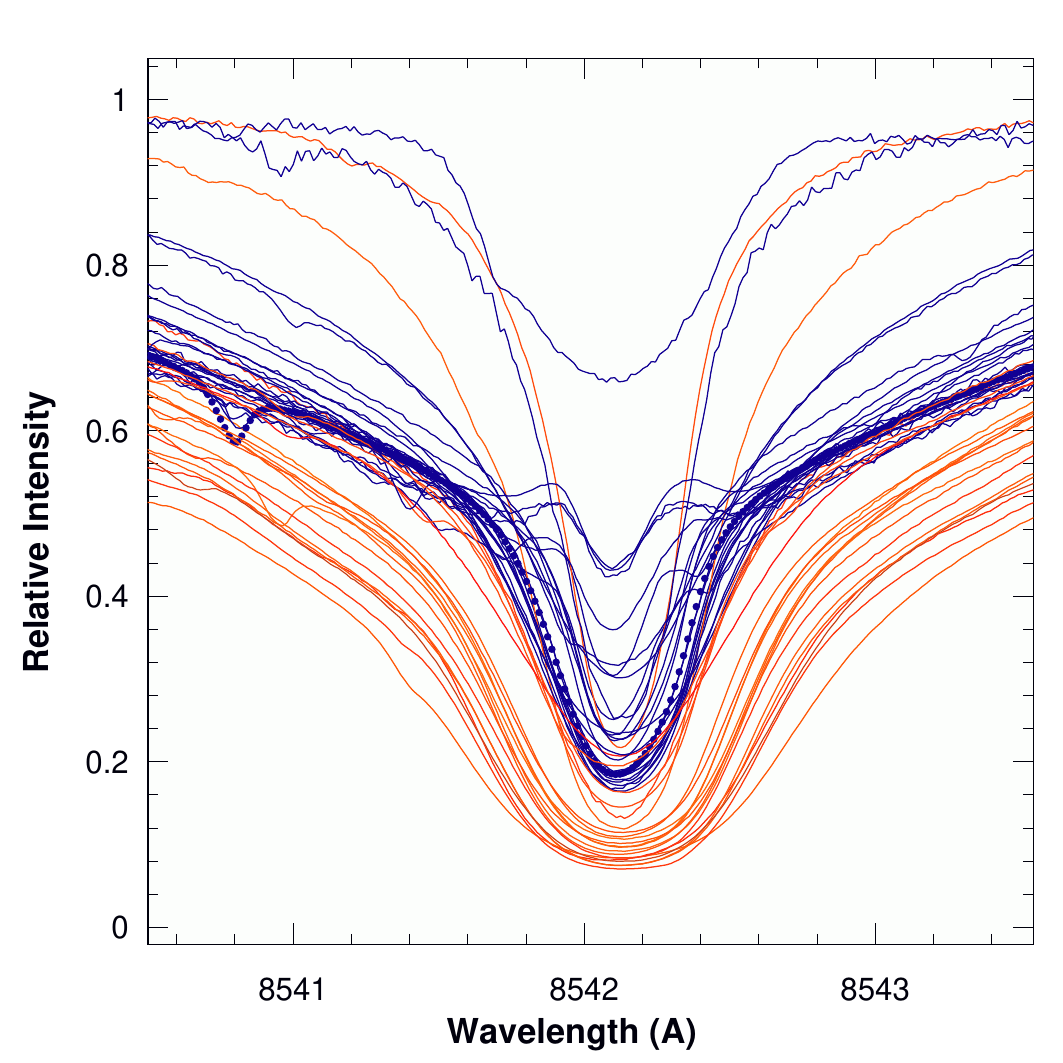}
  \includegraphics[width=.43\textwidth]{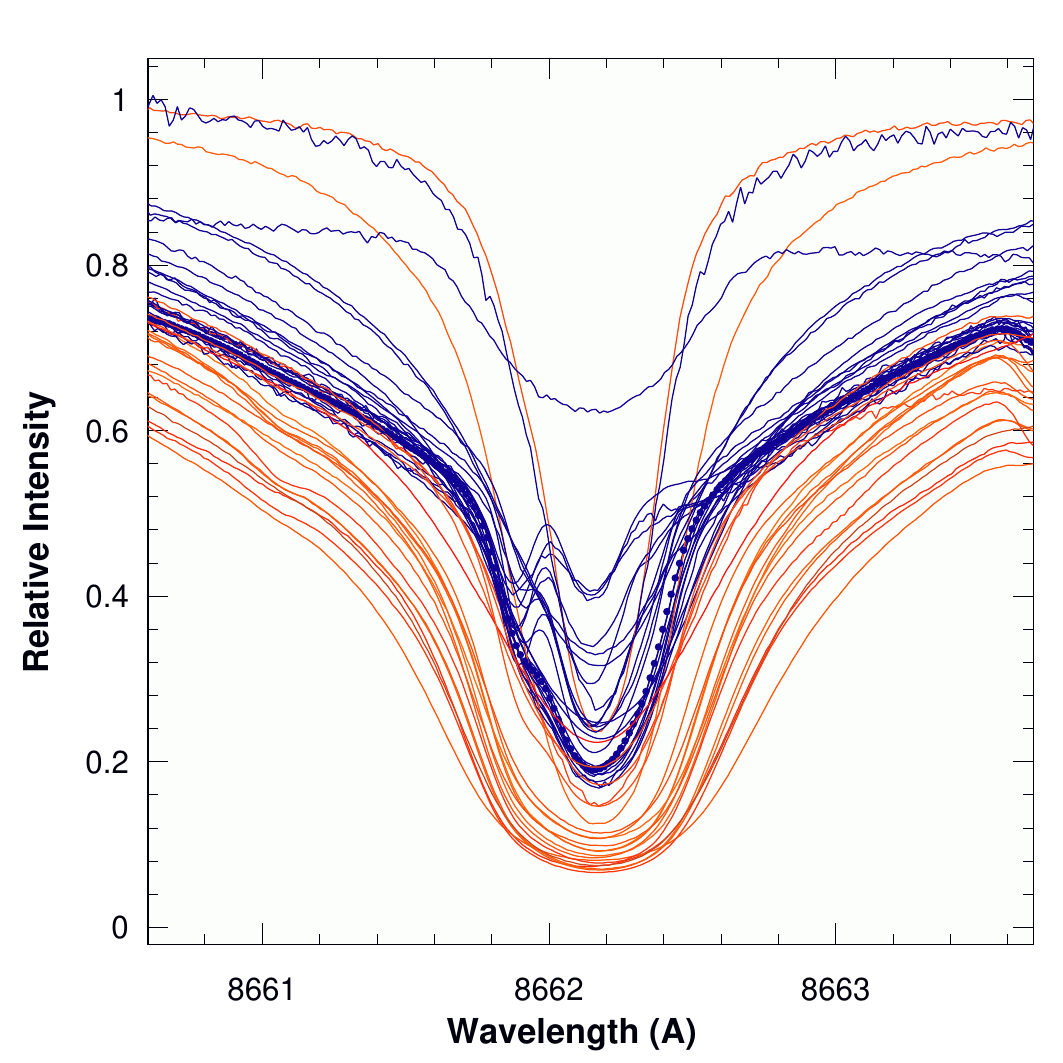}
  \caption{
  Ultra-high resolution \ion{Ca}{ii} IRT-2 (left) and IRT-3 (right) line profiles of the {\it Gaia} benchmark stars observed with PEPSI. Blue spectra are the dwarf stars, red spectra the giants. The solar spectrum is shown as dots. IRT-1 is shown in Fig.~\ref{F-gbscomp} in the main text. 
  }
  \label{F_gbs2-3}
\end{figure*}

\begin{figure*}
\centering
  \includegraphics[width=.9\textwidth]{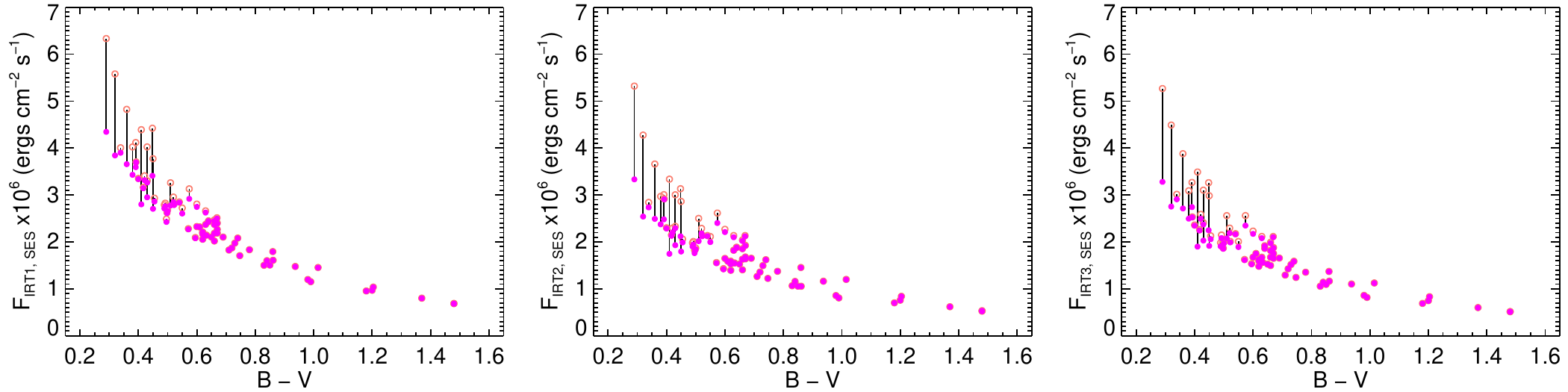}
\caption{
Visualisation of the $v\sin i$ correction using fluxes based on STELLA-SES spectra. The salmon open circles represent stars observed with STELLA before the $v\sin i$ correction was made and magenta filled circles represent the after positions. For clarity, black lines connect the points with large changes. 
}
  \label{Fi:SESfluxvscol}
\end{figure*}

\begin{figure*}
\centering
  \includegraphics[width=.9\columnwidth]{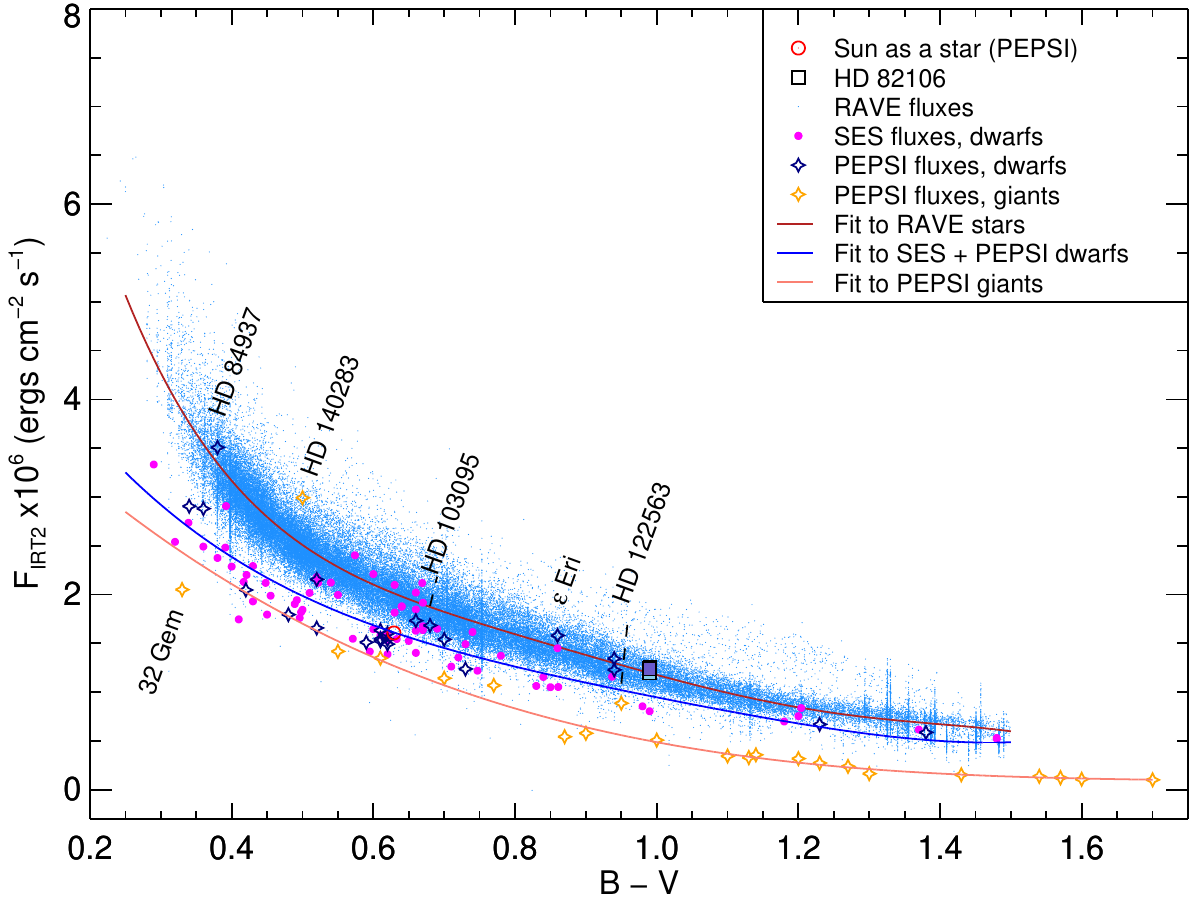}
  \includegraphics[width=.9\columnwidth]{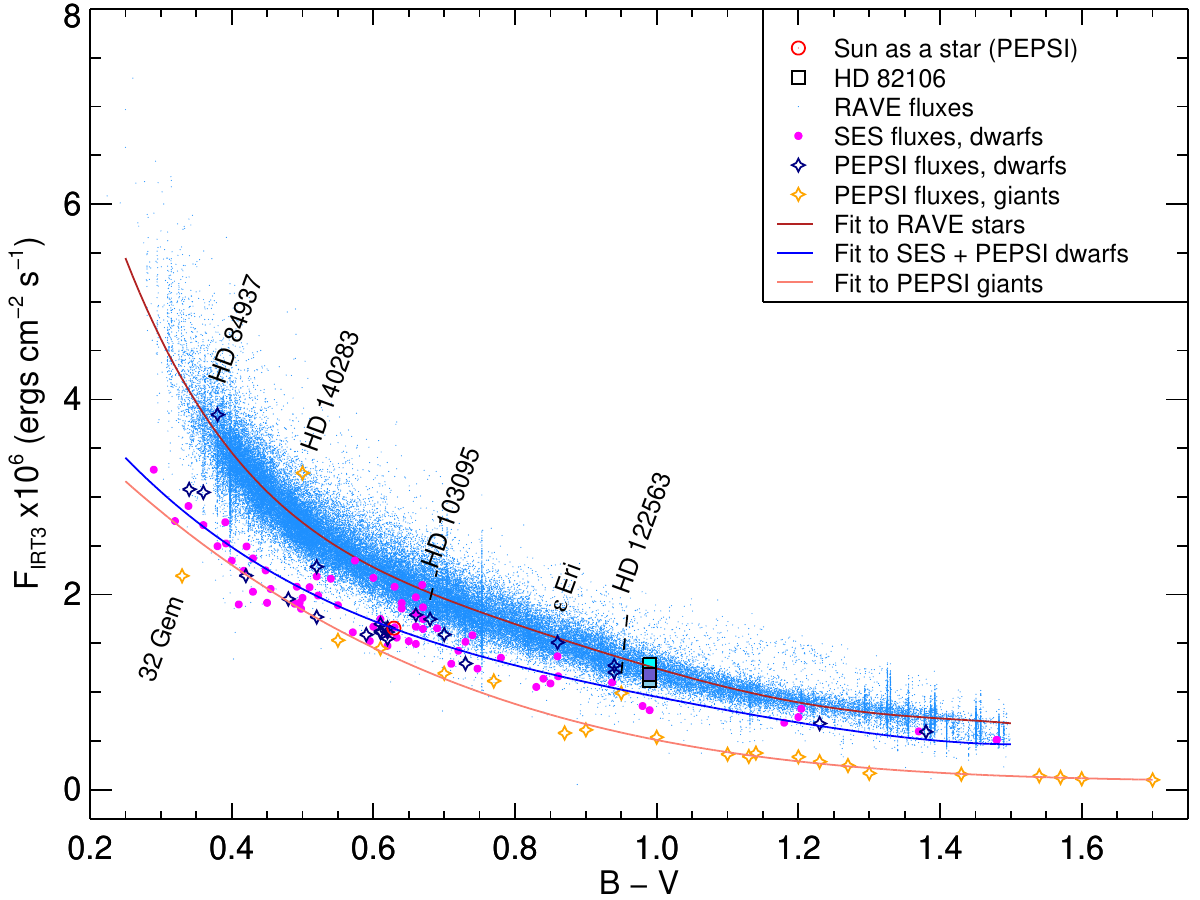}
  \caption{
  As Fig.~\ref{Fallsamples} but for \ion{Ca}{ii} IRT-2 (left) and IRT-3 
(right).
  }
  \label{Fall_IRT2-3}
\end{figure*}

\begin{figure*}
\centering
  \includegraphics[width=.9\columnwidth]{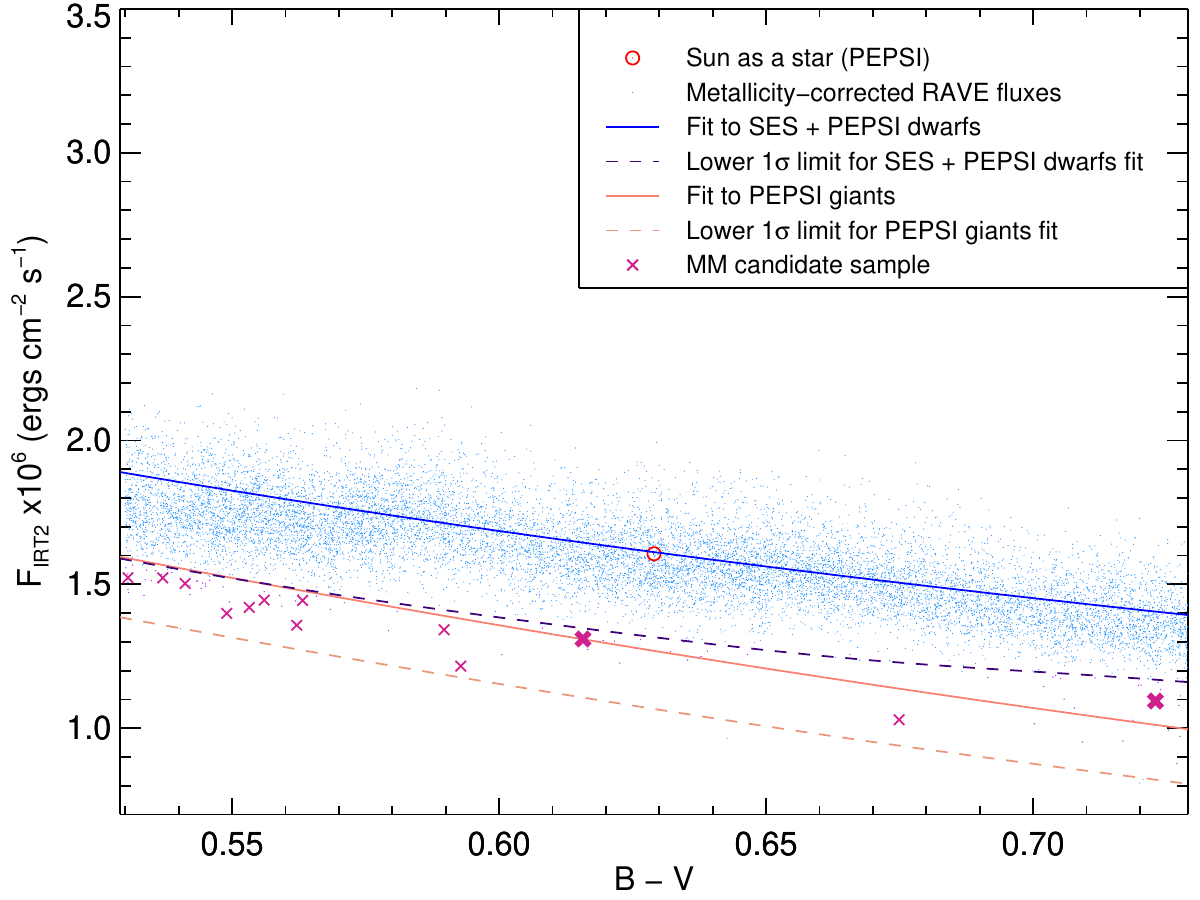}
  \includegraphics[width=.9\columnwidth]{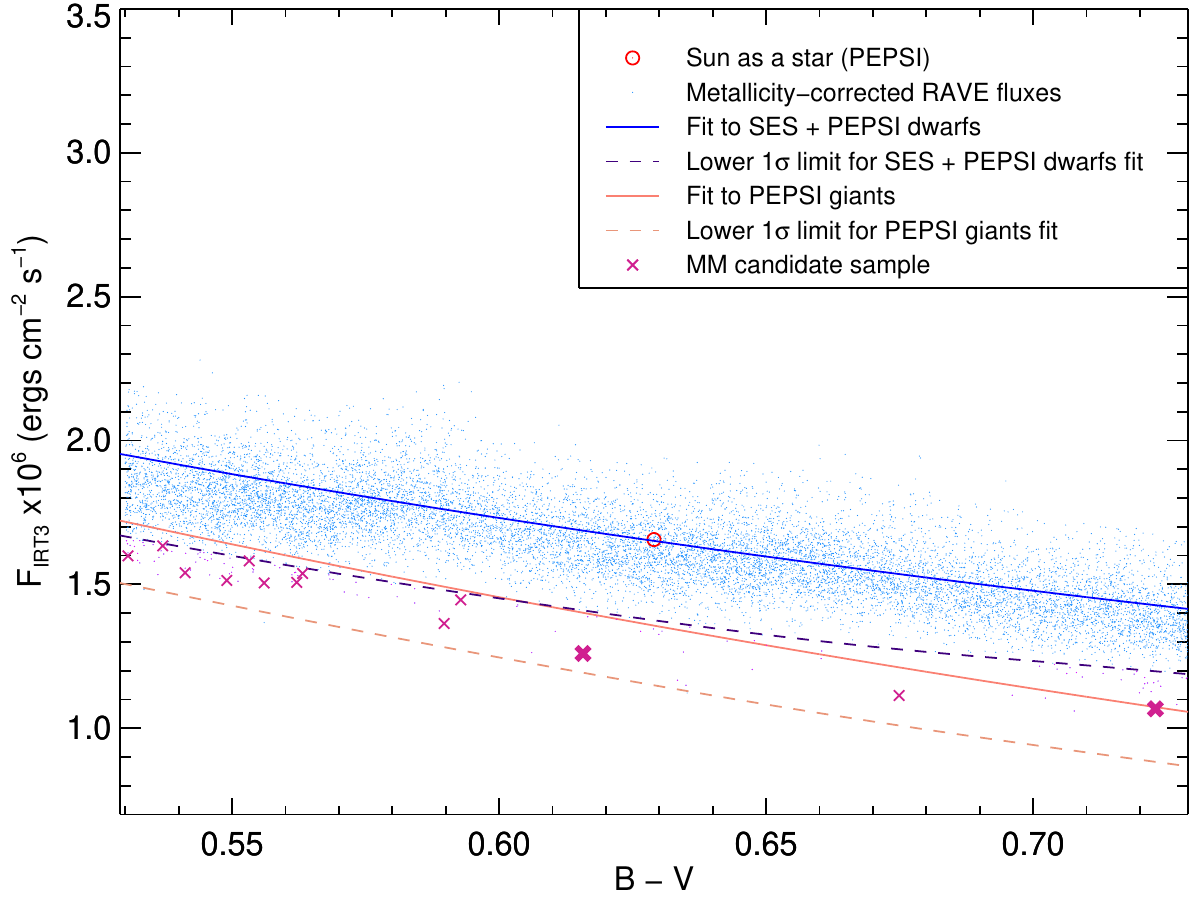}
  \caption{
  As Fig.~\ref{F7} but for \ion{Ca}{ii} IRT-2 (left) and IRT-3 (right).
  }
  \label{metIRT2-3_bin}
\end{figure*}

\begin{figure*}
  \centering
  \includegraphics[width=.9\textwidth]{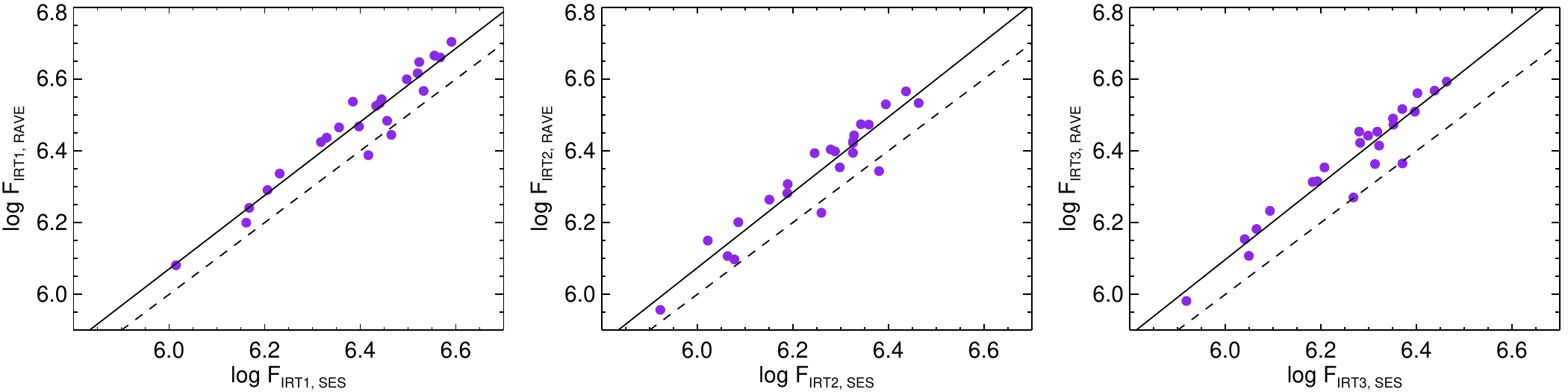}
  \caption{
  STELLA-SES versus RAVE logarithmic fluxes for each of the \ion{Ca}{ii} IRT lines. The dashed black lines represent 1:1 relation while the black solid lines are the fits to the data points.
  }
  \label{Fi:fluxSR}
\end{figure*}

\begin{figure*}
   \centering
   \includegraphics[width=.29\textwidth]{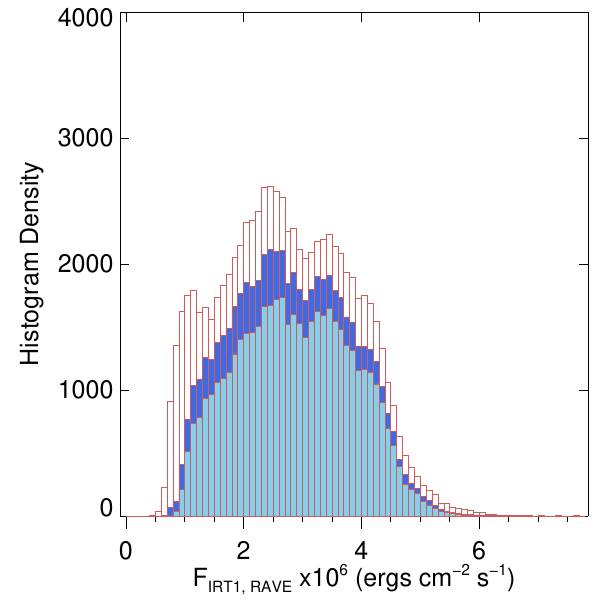}
   \includegraphics[width=.29\textwidth]{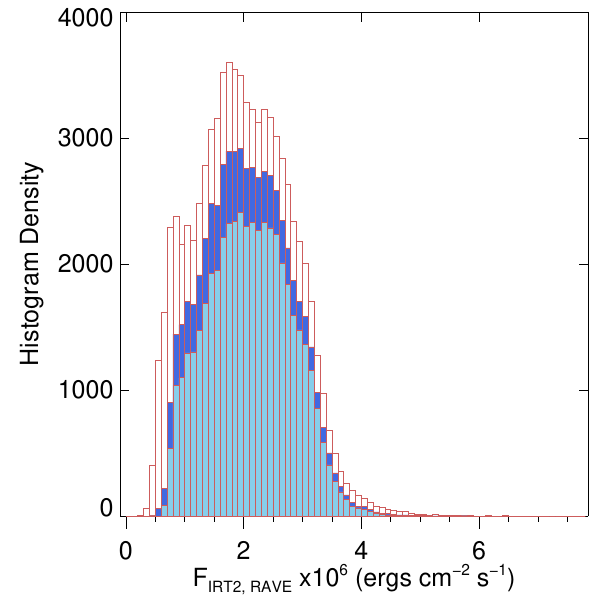}
   \includegraphics[width=.29\textwidth]{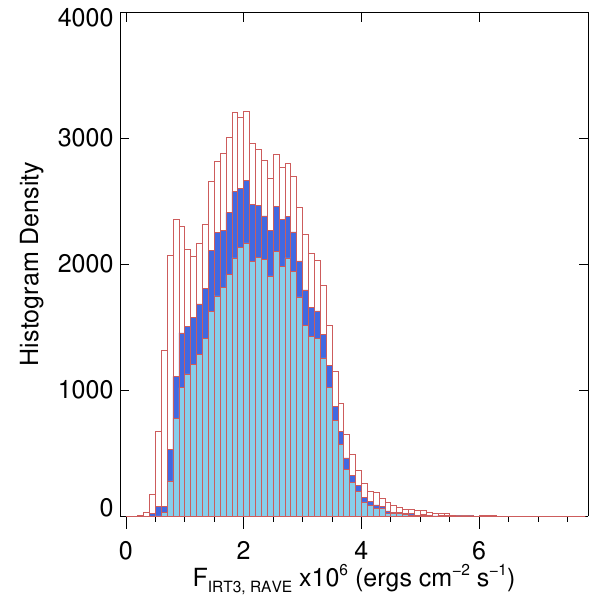}
   \caption{
   Histogram of the (resolution- and metallicity-) uncorrected $F_\mathrm{IRT}$ RAVE fluxes. The three colours indicate increasing S/N for the rejection criterion. White is for the initial 78\,111 RAVE targets with S/N$>$15, dark blue for the sample with S/N$>$30 (58\,672 targets), and light blue for S/N$>$40 (47\,618 targets).
   }
   \label{Rn}
\end{figure*}

\begin{figure}
  \centering
  \includegraphics[width=\columnwidth]{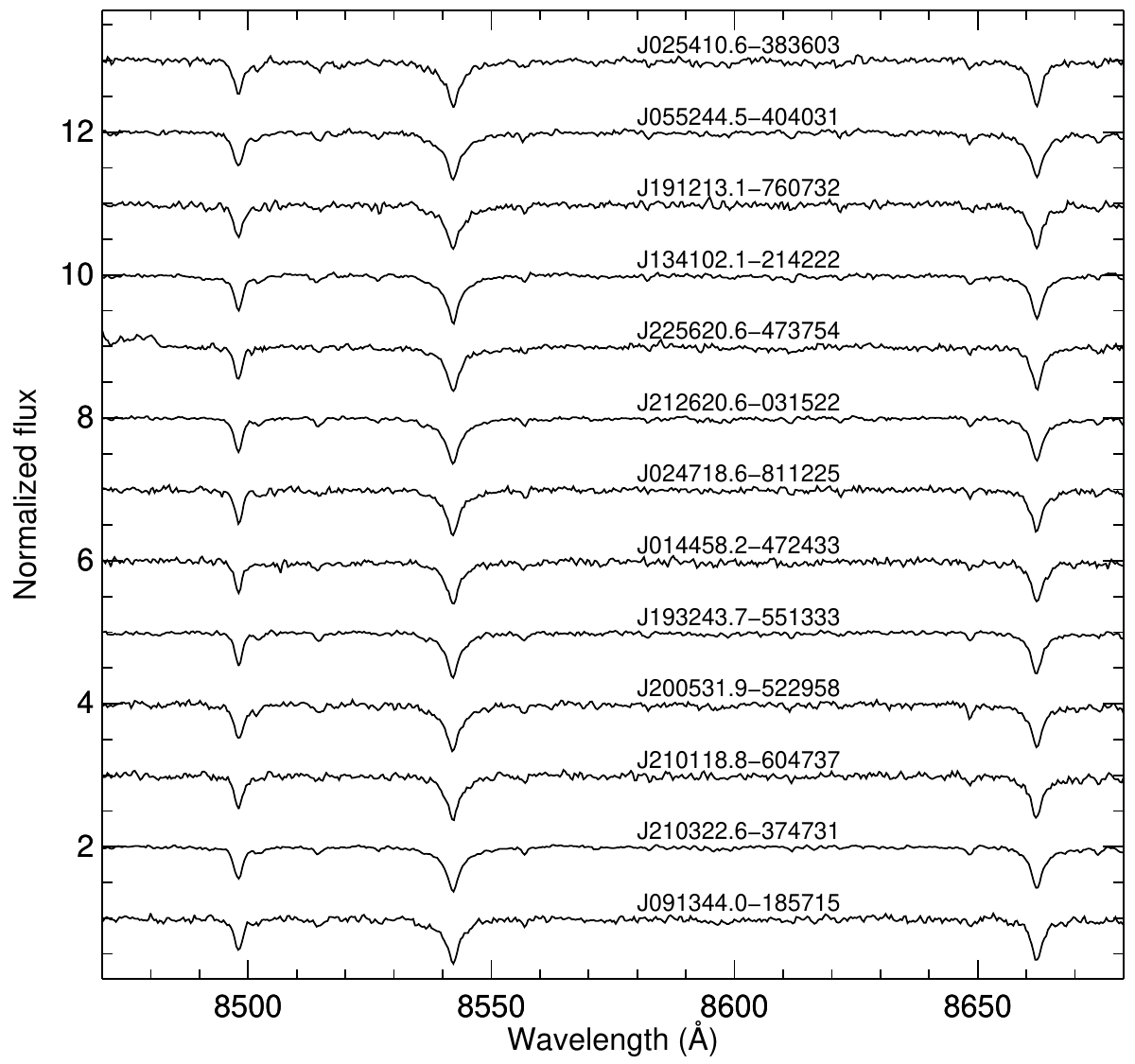}
  \caption{
  RAVE spectra of the 13 lowest-activity targets from Table~\ref{FinalMMcands} including the two MM candidates TIC\,352227373 (RAVE J191213.1-760732) and TYC~7560-477-1 (RAVE J025410.6-383603).
  }
  \label{Fi:MM_13_sample}
\end{figure}

Here we present figures for those IRT lines that were not shown in the main text. Figure~\ref{F_gbs2-3} shows the PEPSI GBS line profiles for IRT-2 and IRT-3. Figure~\ref{Fi:SESfluxvscol} visualises the impact of the $v\sin i$ correction for the STELLA spectra. Figure~\ref{Fall_IRT2-3} shows the uncorrected RAVE fluxes versus B--V for IRT-2 and IRT-3. The same is shown in Fig.~\ref{metIRT2-3_bin} for the corrected RAVE fluxes and limited to a B--V zoom centred around the solar value. 
Figure~\ref{Fi:fluxSR} compares STELLA fluxes with RAVE fluxes for the joint sample and for each of the three \ion{Ca}{ii} IRT lines. The number densities as a function of (uncorrected) fluxes for all RAVE targets and all IRT lines are shown in Fig.~\ref{Rn}. Figure~\ref{Fi:MM_13_sample} plots the observed RAVE spectra for the full sample of 13 MM candidates. 

\FloatBarrier

\subsection{Flux-flux relations}

\begin{figure*}
  \centering
  \includegraphics[width=.85\textwidth]{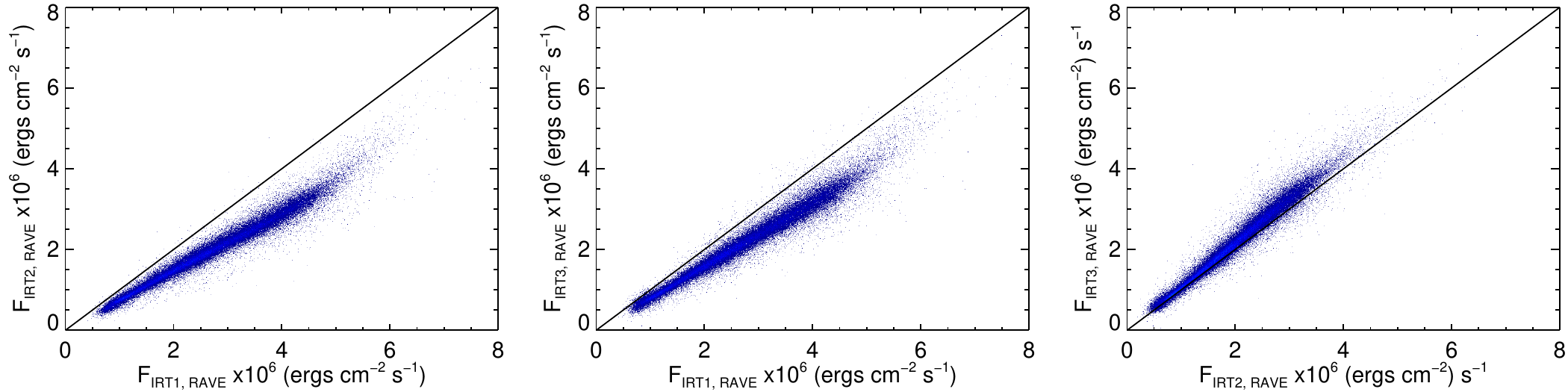}
  \caption{
  Comparison of the obtained line-core fluxes based on RAVE spectra
    for all three \ion{Ca}{ii} IRT lines. The solid lines are 1:1 relations.
    }
  \label{Fi:fluxflux}
\end{figure*}

\begin{figure*}
  \centering
  \includegraphics[width=.85\textwidth]{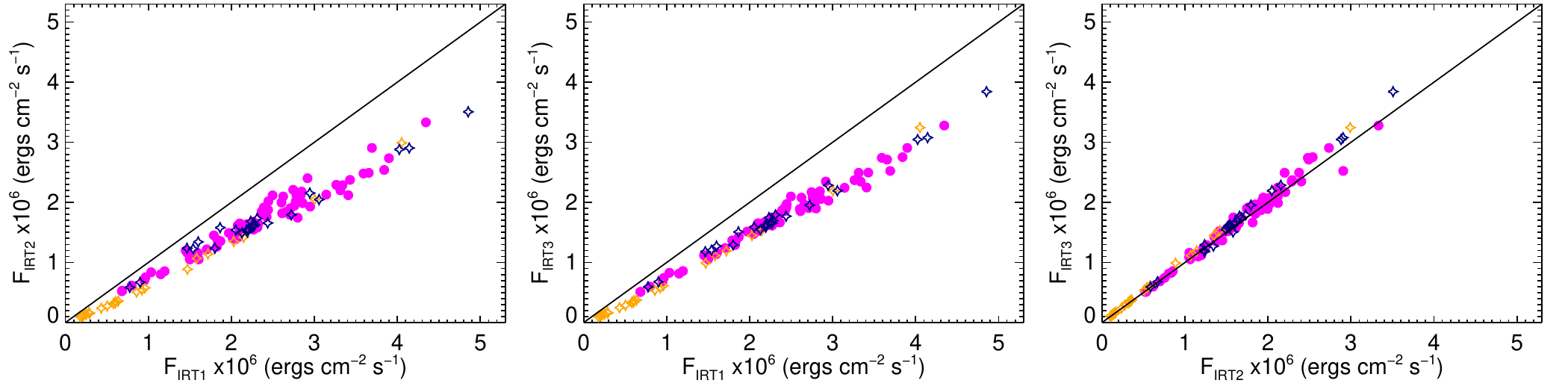}
  \caption{
  As Fig.~\ref{Fi:fluxflux}, but using the fluxes based on STELLA-SES and PEPSI spectra. The symbols are as in Fig.~\ref{Fallsamples}.
  }
  \label{Fi:SESfluxflux}
\end{figure*}

\emph{IRT-IRT.} A comparison of the (uncorrected) RAVE fluxes for the three IRT lines is shown in Fig.~\ref{Fi:fluxflux}. The plots verify that the line-core fluxes are tightly correlated with each other. Apart from the highest flux values of IRT-1 the fluxes follow a linear relationship up to $\approx 5 \times 10^6$ \ecms. The standard deviation for IRT-1 vs.\ IRT-2 is 0.30$\times 10^{6}$, for IRT-1 vs.\ IRT-3 0.24$\times 10^{6}$, and for IRT-2 vs.\ IRT-3 0.30$\times 10^{6}$. It is also noted that IRT-2 is the strongest line (lowest photospheric core flux) while IRT-1 is the weakest line (highest core flux). This is also the reason why IRT-2 is more sensitive to chromospheric filling. A similar plot is given in Fig.~\ref{Fi:SESfluxflux} but for the high-resolution STELLA-SES and PEPSI IRT fluxes. This plot is based on the sample of 81 stars from Tables~\ref{BT:Targets} and \ref{BT:MKs}. It shows the same linear relations as for the RAVE fluxes, but with a flux range that does not extend to the high $\approx 5 \times 10^6$ level of the RAVE spectra.

\begin{figure*}
  \centering
  \includegraphics[width=.85\textwidth]{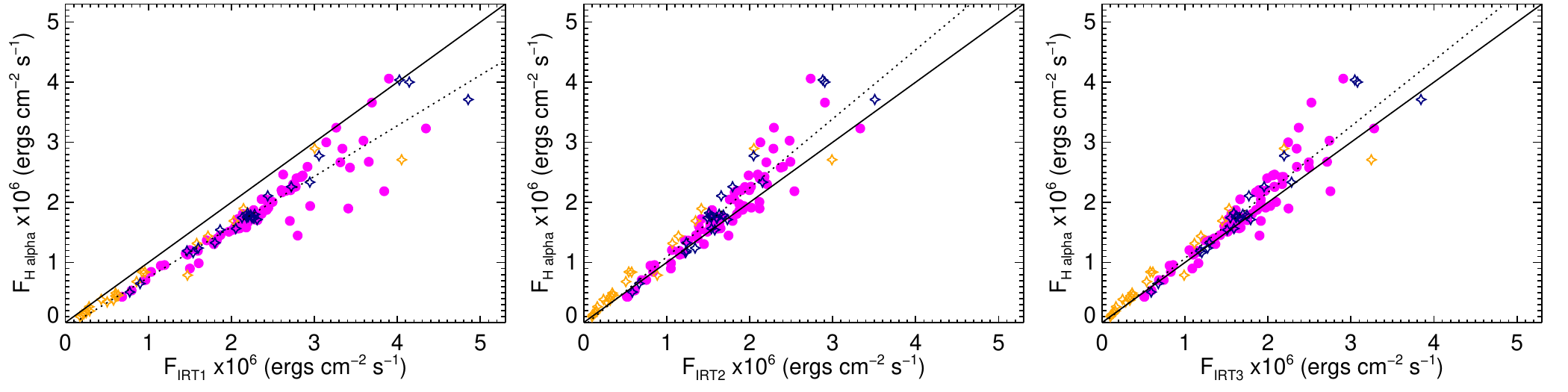}
  \caption{
  Flux-flux relations between \ion{Ca}{ii} IRT lines and \Halpha\ using STELLA-SES and PEPSI spectra. The symbols are as in Fig.~\ref{Fallsamples}. The dashed lines are the fits to the data points.
  }
  \label{Fi:IRTHA}
\end{figure*}

\emph{IRT-others.} STELLA-SES spectra contain not only the \ion{Ca}{ii} IRT region but cover the entire wavelength range from 390 to 870\,nm and thus enable flux-flux relationships with other chromospheric indicators. In Fig.~\ref{Fi:IRTHA} such a relationship is presented for the IRT lines versus the \Halpha\ line in which case also the PEPSI spectra could be included. Simple linear fits, shown in the individual plots, were sufficient to fit the respective slopes within the observational uncertainties. These fits are as follows:
\begin{equation}\label{eqhaIRT1}
  \mathscr{F}_{\mathrm{H\alpha}} = 0.842\mathscr{F}_{\mathrm{IRT-1}} -0.094,
\end{equation}
\begin{equation}\label{eqhaIRT2}
  \mathscr{F}_{\mathrm{H\alpha}} = 1.149\mathscr{F}_{\mathrm{IRT-2}} -0.058,
\end{equation}
\begin{equation}\label{eqhaIRT3}
  \mathscr{F}_{\mathrm{H\alpha}} = 1.098\mathscr{F}_{\mathrm{IRT-3}} -0.028.
\end{equation}

The same procedure is applied to the \ion{Ca}{ii} H\&K lines. Fig.~\ref{Fi:IRTCaHK} in the Appendix shows the relationship between fluxes from IRT and H\&K. The fits are again as follows:
\begin{equation}\label{eqCaHIRT1}
\mathscr{F}_{\mathrm{\ion{Ca}{ii}H}} = 1.127\mathscr{F}_{\mathrm{IRT-1}} - 1.651,
\end{equation}
\begin{equation}\label{eqCaHIRT2}
\mathscr{F}_{\mathrm{\ion{Ca}{ii}H}} = 1.431\mathscr{F}_{\mathrm{IRT-2}} - 1.498,
\end{equation}
\begin{equation}\label{eqCaHIRT3}
\mathscr{F}_{\mathrm{\ion{Ca}{ii}H}} = 1.399\mathscr{F}_{\mathrm{IRT-3}} - 1.477,
\end{equation}
\begin{equation}\label{eqCaKIRT1}
\mathscr{F}_{\mathrm{\ion{Ca}{ii}K}} = 1.127\mathscr{F}_{\mathrm{IRT-1}} - 1.546,
\end{equation}
\begin{equation}\label{eqCaKIRT2}
\mathscr{F}_{\mathrm{\ion{Ca}{ii}K}} = 1.508\mathscr{F}_{\mathrm{IRT-2}} - 1.634,
\end{equation}
\begin{equation}\label{eqCaKIRT3}
\mathscr{F}_{\mathrm{\ion{Ca}{ii}K}} = 1.428\mathscr{F}_{\mathrm{IRT-3}} - 1.535.
\end{equation}
The instrumental scatter is considerably larger in H\&K than in the IRT, mostly because of the low S/N of the STELLA spectra for the blue wavelengths. Fit qualities, in terms of rms, are therefore 0.25 for Eqs.~\ref{eqhaIRT1}--\ref{eqhaIRT3} but only 0.35 for Eqs.~\ref{eqCaHIRT1}--\ref{eqCaHIRT3} and 0.45 for Eqs.~\ref{eqCaKIRT1}--\ref{eqCaKIRT3}. The respective slopes between IRT vs. \Halpha\ and IRT vs. H\&K are clearly different though. The IRT--H\&K slope appears approximately a factor two steeper than the IRT--\Halpha\ slope which is not surprising due the significantly lower photospheric-to-chromospheric contrast in \Halpha\ compared to \ion{Ca}{ii} H\&K.

\begin{figure*}
\centering
  \includegraphics[width=.85\textwidth]{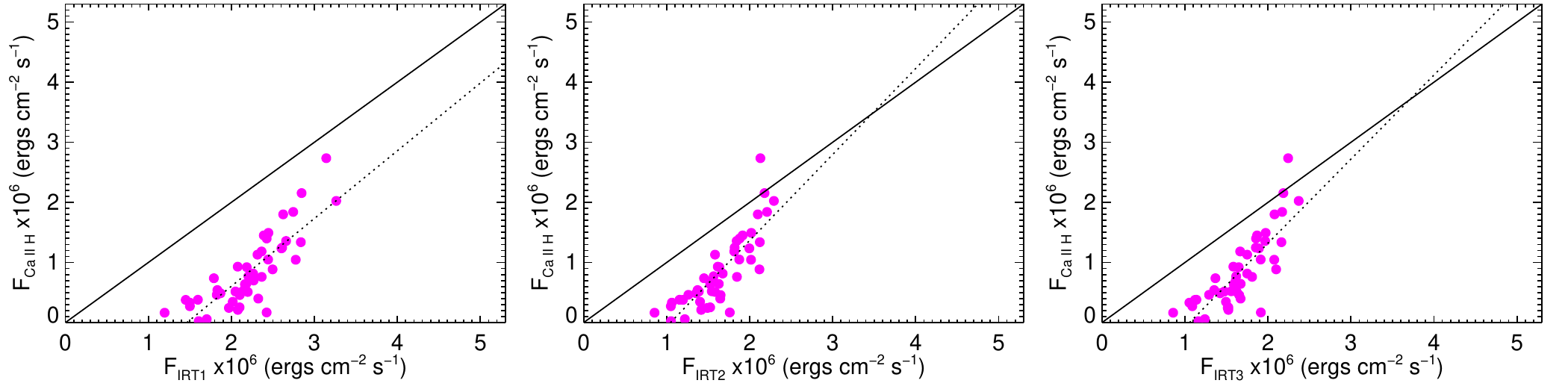}
\includegraphics[width=.85\textwidth]{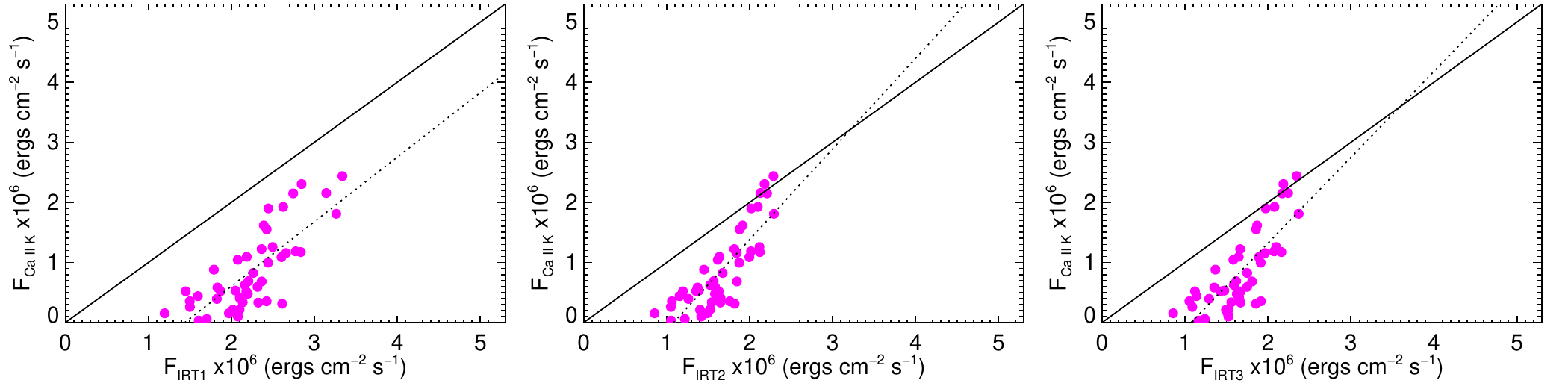}
  \caption{Flux-flux relationships between \ion{Ca}{ii} IRT lines and
    \ion{Ca}{ii}~H\&K line using STELLA-SES spectra. The solid black lines represent
    1:1 relations while the dashed lines are the fits to the data points.
    \emph{Top:} IRT lines vs. \ion{Ca}{ii}~H.
    \emph{Bottom:} IRT lines vs. \ion{Ca}{ii}~K.
  }
  \label{Fi:IRTCaHK}
\end{figure*}


\FloatBarrier
\twocolumn

\begin{sidewaystable*}
\section{Target tables}

Targets jointly observed by RAVE and STELLA are listed in Table~\ref{BT:Targets}. The M-K standards observed with STELLA-SES are listed in Table~\ref{BT:MKs}. Table~\ref{BT:gbs} lists the northern-hemisphere {\it Gaia} benchmark stars observed with PEPSI.

Table~\ref{TappMMcands} lists initial sample of MM candidate stars from the RAVE survey before metallicity correction was applied. The relevant parameters for the metallicity correction are given in Table~\ref{TAppMetal}.\\

  \caption{
  Table listing 23 targets that have been observed with both RAVE and STELLA. 
  }
\label{BT:Targets}
  \centering
\begin{small}
\begin{tabular}{lllccclcccllllll}
\hline\hline \noalign{\smallskip}
RAVE ID & Name & M-K    & $T_{\rm eff}$ & $\log g$  & [Fe/H] & $v\sin i$ & (B--V)$_0$ & N$_R$& N$_S$ 
& $\mathscr{F}_{\mathrm{IRT-1}}^{\mathrm{RAVE}}$ & $\mathscr{F}_{\mathrm{IRT-2}}^{\mathrm{RAVE}}$ & $\mathscr{F}_{\mathrm{IRT-3}}^{\mathrm{RAVE}}$ & $\mathscr{F}_{\mathrm{IRT-1}}^{\mathrm{SES}}$ & $\mathscr{F}_{\mathrm{IRT-2}}^{\mathrm{SES}}$ & $\mathscr{F}_{\mathrm{IRT-3}}^{\mathrm{SES}}$  \\
        &      & class  &  (K)          & (dex)     & (dex)  & (\kms)     & (mag)   &  & & \multicolumn{3}{c}{($10^6$ \ecms)}  & \multicolumn{3}{c}{($10^6$ \ecms)} \\
\noalign{\smallskip}\hline \noalign{\smallskip}
J132119.8-074210 & HD\,116058 & F0~V    & 6710 & 4.03 & $-$0.32 & 48  & 0.45 & 1 & 1 & 3.6886 & 2.6428 & 2.9713 & 3.4089 & 1.1965 & 1.1204 \\
J134539.4-050833 & HD\,119828 & F0~V    & 6728 & 4.27 & $-$0.39 & 4.0 & 0.40 & 1 & 1 & 4.4400 & 2.9723 & 3.2843 & 3.3382 & 2.2862 & 2.3459 \\
J132450.7-102529 & HD\,116611 & F0/2~V  & 6997 & 4.52 & $-$0.16 & 11  & 0.34 & 2 & 1 & 5.0552 & 3.6784 & 3.9153 & 3.8987 & 2.7353 & 2.9070 \\
J134017.2-080110 & HD\,118940 & F2~V    & 6580 & 4.27 & $-$0.16 & 3.2 & 0.39 & 1 & 2 & 4.5721 & 3.4131 & 3.6363 & 3.6936 & 2.9068 & 2.5233 \\
J163201.0-094145 & HD\,148967 & F2/3~V  & 6453 & 4.16 & $-$0.20 & 4.5 & 0.42 & 1 & 2 & 3.9755 & 2.7739 & 3.0900 & 3.1438 & 2.1295 & 2.2432 \\
J160803.7-130011 & HD\,144784 & F3~V    & 6511 & 4.20 & $-$0.34 & 29  & 0.39 & 5 & 2 & 4.6283 & 3.3849 & 3.6957 & 3.5920 & 2.4812 & 2.7414 \\
J140024.0-141320 & HD\,122170 & F3/5~V  & 6446 & 4.08 & $-$0.61 & 11  & 0.42 & 1 & 1 & 4.1330 & 2.9795 & 3.2298 & 3.3136 & 2.2008 & 2.4928 \\
J151252.4-134318 & HD\,134924 & F5~V    & 6423 & 4.24 & $+$0.10 & 9.8 & 0.46 & 1 & 1 & 3.0462 & 2.2589 & 2.3080 & 2.8596 & 1.9880 & 2.0573 \\
J153600.6-185909 & HD\,138971 & F6/7~V  & 6240 & 4.32 & $+$0.02 & 5.2 & 0.52 & 1 & 1 & 3.4972 & 2.6714 & 2.7693 & 2.7848 & 2.1206 & 1.9897 \\
J134558.0-062711 & HD\,119868 & F7~V    & 6258 & 4.22 & $-$0.16 & 9.6 & 0.49 & 5 & 2 & 3.4047 & 2.5031 & 2.8407 & 2.7593 & 1.9427 & 2.0802 \\
J141807.5-095342 & HD\,125195 & F7/8~V  & 6203 & 4.18 & $-$0.15 & 9.4 & 0.50 & 1 & 1 & 3.4431 & 2.4743 & 2.6444 & 2.4250 & 1.7607 & 1.9148 \\
J144052.6-203336 & HD\,129027 & F8/G0~V & 6294 & 4.55 & $+$0.11 & 10.5 & 0.49 & 1 & 1 & 3.3517 & 2.5338 & 2.8413 & 2.7124 & 1.9028 & 1.9055 \\
J191041.8-294849 & HD\,178673 & G0~V    & 6227 & 4.03 & $+$0.04 & 4.2 & 0.50 & 1 & 1 & 2.4427 & 1.6881 & 1.8625 & 2.6122 & 1.8191 & 1.8541 \\
J141310.2-240529 & HD\,124279 & G0/2~V  & 5638 & 4.54 & $+$0.01 & 5.3 & 0.67 & 1 & 1 & 2.9371 & 2.4807 & 2.5976 & 2.4970 & 2.1180 & 2.0983 \\
J190106.0-284250 & HD\,176367 & G1~V    & 6030 & 4.13 & $+$0.11 & 20.0 & 0.57 & 1 & 1 & 2.7821 & 2.2055 & 2.3162 & 2.9176 & 2.4021 & 2.3478 \\
J144225.2-263451 & HD\,129274 & G2~V    & 5974 & 4.02 & $+$0.20 & 4.6 & 0.60 & 1 & 1 & 2.6573 & 1.8373 & 2.0573 & 2.0786 & 1.4161 & 1.5232 \\
J152054.9-182822 & HD\,136367 & G2/3~V  & 6234 & 4.49 & $+$0.06 & 4.3 & 0.57 & 2 & 1 & 2.9183 & 2.0293 & 2.2581 & 2.2687 & 1.5467 & 1.6126 \\
J163053.8-253932 & HD\,148702 & G3~V    & 5805 & 4.51 & $+$0.10 & 3.4 & 0.63 & 1 & 1 & 2.7340 & 1.9135 & 2.0660 & 2.1369 & 1.5439 & 1.5582 \\
J150102.2-021853 & HD\,132775 & G6~V    & 5484 & 4.35 & $+$0.17 & 2.6 & 0.75 & 1 & 1 & 2.1708 & 1.5886 & 1.7089 & 1.7038 & 1.2192 & 1.2402 \\
J141543.7-112907 & HD\,124775 & G8~V    & 4942 & 4.09 & $-$0.54 & 2.2 & 0.86 & 1 & 1 & 1.9531 & 1.4122 & 1.5210 & 1.6070 & 1.0525 & 1.1622 \\
J133106.2-040620 & HIP\,65940 & K2~V    & 4892 & 4.55 & $-$0.27 & 3.6 & 0.94 & 2 & 1 & 1.7424 & 1.2785 & 1.4250 & 1.4715 & 1.1578 & 1.0982 \\
J092954.8+053919 & HD\,82106  & K3~V    & 4731 & 4.84 & $-$0.06 & 3.3 & 0.99 & 1 & 1 & 1.5843 & 1.2514 & 1.2805 & 1.4505 & 1.1965 & 1.1204 \\
J160016.5-014755 & HIP\,78395 & K6~V    & 4489 & 4.93 & $-$0.20 & 3.5 & 1.20 & 4 & 1 & 1.2055 & 0.9056 & 0.9589 & 1.0340 & 0.8372 & 0.8292\\
\hline
\end{tabular}
\end{small}
\tablefoot{The first column has RAVE IDs, the second column gives the names of the targets and the third column has their M-K classes. The fourth, the fifth, the sixth, and the seventh columns give stellar parameters effective temperature, surface gravity, metallicity, and $v\sin i$ value, respectively,  measured from STELLA-SES spectra. The eighth column has the B--V$_0$ colour. The ninth and the tenth columns give the respective number of spectra obtained with RAVE (N$_R$) and STELLA (N$_S$) followed flux values for each of the IRT lines based on RAVE spectra in columns 11--13 and STELLA spectra in columns 14-16.}
\end{sidewaystable*}

\begin{table*}
\caption{
Table listing 49 M-K standard stars observed with STELLA-SES selected from \citet{1989BICDS..36...27G}.
}
\label{BT:MKs}
\begin{tabular}{llccrlclll}
\hline\hline \noalign{\smallskip}
Name & M-K          & $T_{\rm eff}$ & $\log g$  & [Fe/H] & $v\sin i$ & (B--V)$_0$ & $\mathscr{F}_{\mathrm{IRT-1}}$ & $\mathscr{F}_{\mathrm{IRT-2}}$ & $\mathscr{F}_{\mathrm{IRT-3}}$ \\
     &  class       & (K)           & (dex)     & (dex)  & (\kms)    & (mag)  & \multicolumn{3}{c}{($10^6$ \ecms)} \\
\noalign{\smallskip}\hline \noalign{\smallskip}
HD\,23585  & F0~V   & 7440 & 4.29 & $+$0.04 & 87   & 0.29 & 4.3447 & 3.3324 & 3.2791 \\
HD\,58946  & F1~V   & 6709 & 4.13 & $-$0.26 & 65   & 0.32 & 3.8420 & 2.5396 & 2.7529 \\
HD\,26690  & F2~V   & 6065 & 3.18 & $-$0.25 & 48   & 0.36 & 3.6560 & 2.4910 & 2.7119 \\
HD\,26015  & F3~V   & 6720 & 3.97 & $+$0.28 & 31   & 0.38 & 3.4296 & 2.3747 & 2.4953 \\
HD\,26911  & F3~V   & 6839 & 4.15 & $+$0.21 & 70   & 0.41 & 2.7995 & 1.7456 & 1.8984 \\
HD\,134083 & F5~V   & 6237 & 4.17 & $-$0.05 & 49   & 0.43 & 2.9498 & 1.9298 & 2.0283 \\
HD\,27534  & F5~V   & 6545 & 4.05 & $+$0.75 & 50   & 0.45 & 2.7050 & 1.7946 & 1.9150 \\
HD\,210027 & F5~V   & 6419 & 4.15 & $-$0.11 & 6.8  & 0.43 & 3.2637 & 2.2923 & 2.3704 \\
HD\,126660 & F7~V   & 6292 & 4.06 & $-$0.03 & 31   & 0.51 & 2.7757 & 2.0164 & 2.0753 \\
HD\,222368 & F7~V   & 6169 & 4.09 & $-$0.14 & 5.9  & 0.50 & 2.6597 & 1.8438 & 1.9658 \\
HD\,27808  & F8~V   & 6217 & 4.31 & $+$0.01 & 13   & 0.52 & 2.8465 & 2.1794 & 2.1856 \\
HD\,98231  & F8.5~V & 5932 & 4.11 & $-$0.33 & 4.2  & 0.54 & 2.8374 & 2.1226 & 2.1622 \\
HD\,22484  & F9~V   & 6000 & 4.06 & $-$0.08 & 4.0  & 0.85 & 1.5005 & 1.0506 & 1.0886 \\
HD\,27383  & F9~V   & 5912 & 4.11 & $-$0.06 & 14   & 0.55 & 2.6042 & 1.9964 & 1.8905 \\
HD\,13974  & G0~V   & 5860 & 4.41 & $-$0.47 & 3.6  & 0.63 & 2.3643 & 1.8150 & 1.6653 \\
HD\,27836  & G0~V   & 5843 & 4.35 & $-$0.12 & 7.8  & 0.63 & 2.6246 & 2.0984 & 2.0791 \\
HD\,39587  & G0~V   & 5883 & 4.47 & $-$0.03 & 9.9  & 0.60 & 2.7455 & 2.2093 & 2.1710 \\
HD\,109358 & G0~V   & 6013 & 4.41 & $-$0.20 & 2.4  & 0.61 & 2.3152 & 1.5829 & 1.7495 \\
HD\,143761 & G0~V   & 5823 & 4.19 & $-$0.19 & 2.2  & 0.60 & 2.3243 & 1.6458 & 1.6702 \\
HD\,95128  & G1~V   & 5788 & 3.99 & $+$0.03 & 3.3  & 0.62 & 2.1634 & 1.5248 & 1.5894 \\
HD\,10307  & G1.5~V & 5780 & 4.06 & $+$0.03 & 2.6  & 0.62 & 2.2055 & 1.5666 & 1.6174 \\
HD\,20619  & G1.5~V & 5669 & 4.29 & $-$0.22 & 2.7  & 0.66 & 2.1798 & 1.6283 & 1.6710 \\
HD\,30495  & G1.5~V & 5652 & 4.13 & $+$0.03 & 3.4  & 0.64 & 2.4260 & 1.8796 & 1.8564 \\
HD\,45184  & G1.5~V & 5743 & 4.12 & $+$0.04 & 3.2  & 0.62 & 2.1976 & 1.5830 & 1.6247 \\
HD\,84737  & G2~V   & 5940 & 4.05 & $+$0.16 & 3.3  & 0.62 & 2.0495 & 1.3880 & 1.4744 \\
HD\,98230  & G2~V   & 5890 & 4.04 & $-$0.35 & 3.0  & 0.64 & 2.4426 & 1.8750 & 1.9127 \\
HD\,146233 & G2~V   & 5814 & 4.45 & $+$0.04 & 2.7  & 0.65 & 2.0996 & 1.5244 & 1.5224 \\
HD\,28099  & G2~V   & 5750 & 4.29 & $+$0.17 & 3.8  & 0.66 & 2.3656 & 1.8471 & 1.8092 \\
HD\,1835   & G2.5~V & 5735 & 4.30 & $+$0.23 & 5.9  & 0.66 & 2.4464 & 2.0199 & 1.9724 \\
HD\,76151  & G3~V   & 5679 & 4.27 & $+$0.12 & 2.4  & 0.67 & 2.1857 & 1.6356 & 1.6480 \\
HD\,186427 & G3~V   & 5807 & 4.35 & $+$0.07 & 2.3  & 0.66 & 2.0171 & 1.4025 & 1.4940 \\
HD\,117176 & G4~V   & 5473 & 3.94 & $-$0.06 & 2.2  & 0.71 & 1.8248 & 1.2620 & 1.2905 \\
HD\,20630  & G5~V   & 5645 & 4.31 & $+$0.08 & 4.3  & 0.67 & 2.3909 & 1.9158 & 1.8691 \\
HD\,224930 & G5~V   & 5445 & 4.39 & $-$0.79 & 2.2  & 0.67 & 2.2629 & 1.6753 & 1.7497 \\
HD\,101501 & G8~V   & 5310 & 4.42 & $-$0.05 & 3.0  & 0.74 & 2.0769 & 1.6159 & 1.5829 \\
HD\,196761 & G8~V   & 5372 & 4.28 & $-$0.31 & 2.2  & 0.72 & 1.8690 & 1.3551 & 1.4242 \\
HD\,3651   & K0~V   & 5074 & 4.17 & $+$0.18 & 2.4  & 0.83 & 1.4996 & 1.0632 & 1.0533 \\
HD\,165341 & K0~V   & 5184 & 4.30 & $+$0.05 & 3.0  & 0.86 & 1.7898 & 1.4501 & 1.3668 \\
HD\,185144 & K0~V   & 5289 & 4.56 & $-$0.21 & 1.9  & 0.78 & 1.8319 & 1.3726 & 1.3524 \\
HD\,224618 & K0~V   & 5089 & 4.32 & $-$0.76 & 2.7  & 0.73 & 1.9706 & 1.4915 & 1.5151 \\
HD\,6582   & K1~V   & 5178 & 4.07 & $-$0.75 & 2.0  & 0.69 & 2.1010 & 1.6499 & 1.6538 \\
HD\,10476  & K1~V   & 5129 & 4.52 & $-$0.04 & 2.6  & 0.84 & 1.5971 & 1.1551 & 1.1378 \\
HD\,16160  & K3~V   & 4583 & 4.52 & $-$0.13 & 2.6  & 0.98 & 1.1962 & 0.8543 & 0.8578 \\
HD\,219134 & K3~V   & 4800 & 4.55 & $+$0.06 & 2.5  & 0.99 & 1.1496 & 0.8040 & 0.8150 \\
HD\,201091 & K5~V   & 4398 & 4.63 & $-$0.13 & 3.2  & 1.18 & 0.9514 & 0.6987 & 0.6864 \\
HD\,10436  & K5.5~V & 4172 & 4.48 & $-$0.15 & 3.6  & 1.20 & 0.9664 & 0.7543 & 0.7441 \\
HD\,201092 & K7~V   & 4174 & 4.68 & $-$0.21 & 3.2  & 1.37 & 0.7987 & 0.6161 & 0.5976 \\
HD\,42581  & M1~V   & 3779 & 4.28 & $-$0.10 & 4.2  & 1.48 & 0.6817 & 0.5215 & 0.5113 \\
HD\,36395  & M1.5~V & 3771 & 4.36 & $+$0.21 & 4.2  & 1.48 & 0.6843 & 0.5309 & 0.5099 \\
\hline
\end{tabular}
\tablefoot{The first column has the names of the targets, the second column the spectral types, followed by stellar parameters effective temperature, surface gravity, metallicity, and $v\sin i$ values in columns 3--6. The seventh columns gives the (B--V)$_0$ colour followed by flux values for each of the IRT lines in columns 8--10.}
\end{table*}

\begin{table*}
\caption{
IRT fluxes from PEPSI spectra.}\label{BT:gbs}
\begin{tabular}{llllllllll}
\hline\hline \noalign{\smallskip}
Star & M-K   &  $T_{\rm eff}$ & $\log g$ & [Fe/H] & $v\sin i$ & (B--V)$_0$ & $\mathscr{F}_{\mathrm{IRT-1}}$ & $\mathscr{F}_{\mathrm{IRT-2}}$ & $\mathscr{F}_{\mathrm{IRT-3}}$ \\
     & class &  (K)           & (dex)    & (dex)  & (\kms)    & (mag)     & \multicolumn{3}{c}{($10^6$ \ecms)}     \\
\noalign{\smallskip}\hline \noalign{\smallskip}
\emph{Giants} & & & & & & & & &   \\
\noalign{\smallskip}
\object{32 Gem}          & A9 III  & 7240 & 1.45 & --0.28 & 5.0 & 0.33  & 3.4905 & 2.3904 & 2.5544 \\
\object{HD 140283}       & F3 IV   & 5788 & 3.75 & --2.48 & 6.0 & 0.50 & 4.1208 & 3.0600 & 3.3142 \\
\object{$\eta$ Boo}      & G0 IV   & 6099 & 3.94 & +0.26 & 11  & 0.55 & 2.2494 & 1.5045 & 1.6306 \\
\object{$\zeta$ Her}     & G0 IV   & 5759 & 3.69 & +0.02 & 3.1 & 0.61 & 2.0750 & 1.3677 & 1.4829 \\
\object{$\delta$ CrB}    & G3.5 III& 5362 & 3.29 & --0.12 & 5.2 & 0.77 & 1.6175 & 1.0950 & 1.1428 \\
\object{$\mu$ Her}       & G5 IV   & 5633 & 4.02 & +0.31 & 2   & 0.70 & 1.7522 & 1.1614 & 1.2252 \\
\object{$\beta$ Boo}     & G8 III  & 5100 & 2.8  & --0.1  & 1   & 0.87 & 0.9447 & 0.5494 & 0.5970 \\
\object{$\epsilon$ Vir}  & G8 III  & 5060 & 2.97 & +0.15 & 2   & 0.90 & 0.9813 & 0.5855 & 0.6312 \\
\object{HD 122563}       & G8III   & 4642 & 1.32 & --2.67 & 7.0 & 0.95 & 1.5216 & 0.9204 & 1.0394 \\
\object{$\beta$ Gem}     & K0 IIIb & 4810 & 2.55 & +0.02 & 3   & 1.00 & 0.8799 & 0.5218 & 0.5588 \\
\object{HD 107328}       & K0 IIIb & 4508 & 2.3  & --0.48 & 2   & 1.13 & 0.6167 & 0.3365 & 0.3511 \\
\object{$\alpha$ UMa}    & K0 III  & 4660 & 2.46 & --0.20 & 6   & 1.10 & 0.6325 & 0.3613 & 0.3816 \\
\object{$\alpha$ Ari}    & K1 IIIb & 4498 & 2.4  & --0.25 & 4.2 & 1.14 & 0.6584 & 0.3721 & 0.3931 \\
\object{$\alpha$ Boo}    & K1.5 III& 4308 & 1.7  & --0.55 & 4.2 & 1.23 & 0.5253 & 0.2851 & 0.3013 \\
\object{7 Psc}           & K2 III  & 4351 & 1.73 & --0.70 & 7.0 & 1.20 & 0.6126 & 0.3367 & 0.3560 \\
\object{$\mu$ Leo}       & K2 III  & 4519 & 2.43 & +0.27 & 4.5 & 1.27 & 0.4468 & 0.2489 & 0.2597 \\
\object{$\gamma$ Aql}    & K3 II   & 4210 & 1.63 & --0.29 & 8   & 1.30 & 0.3083 & 0.1802 & 0.1855 \\
\object{$\beta$ UMi}     & K4 III  & 4030 & 1.83 & --0.29 & 8   & 1.43 & 0.3124 & 0.1678 & 0.1756 \\
\object{$\alpha$ Tau}    & K5 III  & 3921 & 1.17 & --0.20 & 3.5 & 1.54 & 0.2607 & 0.1435 & 0.1488 \\
\object{$\mu$ UMa}       & M0 III  & 3899 & 1.0  & 0.00  & 7.5 & 1.57 & 0.2370 & 0.1333 & 0.1367 \\
\object{$\gamma$ Sge}    & M0 III  & 3904 & 1.06 & --0.26 & 6   & 1.60 & 0.2124 & 0.1159 & 0.1216 \\
\object{$\alpha$ Cet}    & M1.5 III& 3738 & 0.66 & --0.24 & 7   & 1.70 & 0.1881 & 0.1095 & 0.1100 \\
\noalign{\smallskip}
\emph{Dwarfs} & & & & & & & & &   \\
\noalign{\smallskip}
\object{$\alpha$ CMa}    & A1 V    & 9850 & 4.3  & +0.40 & 16.7& 0.00 & 9.9535 & 9.0056 & 8.3011 \\
\object{$\sigma$ Boo}    & F4 V    & 6756 & 4.28 & --0.36 & 7.3 & 0.34 & 4.2224 & 2.9811 & 3.1723 \\
\object{HD 49933}        & F5 V-IV & 6628 & 4.19 & --0.39 & 5   & 0.36 & 4.0838 & 2.9441 & 3.0992 \\
\object{$\alpha$ CMi}    & F5 V-IV & 6582 & 3.98 & --0.02 & 5.4 & 0.42& 3.1220 & 2.0978 & 2.2249 \\
\object{$\theta$ UMa}    & F7 V    & 6217 & 3.83 & --0.17 & 8.8 & 0.48 & 2.8152 & 1.8675 & 2.0413 \\
\object{HD 84937}        & F8 V    & 6356 & 4.1  & --2.0  & 5   & 0.38 & 4.9229 & 3.5873 & 3.9031 \\
\object{$\beta$ Vir}     & F9 V    & 6093 & 4.08 & +0.13 & 3.4 & 0.52 & 2.4794 & 1.6909 & 1.8066 \\
\object{HD 22879}        & F9 V    & 5868 & 4.45 & --0.80 & 3.0 & 0.52 & 2.9752 & 2.1878 & 2.3040 \\
\object{HD 159222}       & G1 V    & 5856 & 4.39 & +0.16 & 2   & 0.61 & 2.2225 & 1.5620 & 1.6471 \\
\object{16 Cyg A}        & G1.5 V  & 5849 & 4.28 & +0.08 & 5   & 0.61 & 2.2240 & 1.5606 & 1.6462 \\
\object{HD 101364}       & G2 V    & 5771 & 4.34 & --0.01 & 1   & 0.62 & 2.2710 & 1.6254 & 1.6832 \\
\object{HD 82943}        & G2 V    & 5918 & 4.34 & +0.23 & 1.4 & 0.59 & 2.2340 & 1.5333 & 1.6082 \\
\object{18 Sco}          & G2 V    & 5824 & 4.42 & +0.03 & 2.1 & 0.61 & 2.3081 & 1.6577 & 1.7123 \\
Sun (11/2016)            & G2 V    & 5771 & 4.44 & 0.00  & 2   & 0.63 & 2.3047 & 1.6363 & 1.6862 \\
\object{51 Peg}          & G2.5 V  & 5758 & 4.32 & +0.18 & 2.0 & 0.62 & 2.1585 & 1.5179 & 1.5758 \\
\object{16 Cyg B}        & G3 V    & 5807 & 4.35 & +0.07 & 2.3 & 0.62 & 2.2205 & 1.5726 & 1.6386 \\
\object{70 Vir}          & G4 V-IV & 5473 & 3.94 & --0.06 & 1   & 0.73 & 1.8243 & 1.2563 & 1.3129 \\
\object{$\mu$ Cas}       & G5 V    & 5358 & 4.49 & --0.83 & 4.1 & 0.66 & 2.3501 & 1.7607 & 1.8201 \\
\object{HD 103095}       & G8 V    & 5235 & 4.72 & --1.33 & 3.1 & 0.68 & 2.2667 & 1.7027 & 1.7640 \\
\object{$\tau$ Cet}      & G8.5 V  & 5463 & 4.52 & --0.51 & 3   & 0.70 & 2.0853 & 1.5612 & 1.6198 \\
\object{$\epsilon$ Eri}  & K2 V    & 5130 & 4.63 & --0.08 & 4   & 0.86 & 1.8753 & 1.5881 & 1.5170 \\
\object{HD 192263}       & K2 V    & 4975 & 4.61 & +0.04 & 3.3 & 0.94 & 1.5477 & 1.2374 & 1.2102 \\
\object{HD 128311}       & K3 V    & 4967 & 4.67 & +0.16 & 3.2 & 0.94 & 1.6079 & 1.3509 & 1.2797 \\
\object{HD 82106}        & K3 V    & 4861 & 4.59 & --0.02 & 3.0 & 0.99& 1.4735 & 1.2390 & 1.1824 \\
\object{61 Cyg A}        & K5 V    & 4398 & 4.63 & --0.13 & 1   & 1.23 & 0.9089 & 0.6748 & 0.6819 \\
\object{61 Cyg B}        & K7 V    & 4174 & 4.68 & --0.21 & 3   & 1.38 & 0.7832 & 0.5904 & 0.5982 \\
\noalign{\smallskip}\hline
\end{tabular}
\tablefoot{The first column gives the name for the target, listing first the giants and then the dwarfs, followed by the spectral class on the second column. Columns 3--6 list the stellar parameters effective temperature, surface gravity, metallicity, and  $v\sin i$ value, respectively. The seventh column has the (B--V)$_0$ colour. The flux values for each of the IRT lines are given in columns 8--10.}
\end{table*}

\begin{table*}
\caption{
Initial temporary (fake) sample of MM candidate stars from the RAVE survey.
}\label{TappMMcands}
\begin{small}
\begin{tabular}{lccrcllll}
\hline\hline \noalign{\smallskip}
RAVE name & $T_{\rm eff}$ & $\log g$  & [Fe/H] & (B--V)$_0$ & $\mathscr{F}_{\mathrm{IRT-1}}$ & $\mathscr{F}_{\mathrm{IRT-2}}$ & $\mathscr{F}_{\mathrm{IRT-3}}$ & $\log R_{\rm IRT}$ \\
     & (K)          & (dex)     & (dex)  & (mag)  & \multicolumn{3}{c}{($10^6$ \ecms)} & ($\sigma T_{\rm eff}$) \\
\noalign{\smallskip}\hline \noalign{\smallskip}
J141417.8-230534 &     6055 &  4.46 &  $+$0.28 &  0.60 & 2.0845 &  1.4245 &   1.4885 & $-$4.18 \\
J125650.6-062817 &     5563 &  4.11 &  $+$0.27 &  0.73 & 1.6468 &  1.1355 &   1.1871 & $-$4.14 \\
J201838.1-351034 &     5672 &  4.19 &  $+$0.37 &  0.70 & 1.6448 &  1.1247 &   1.1260 & $-$4.18 \\
J000839.6-405632 &     5673 &  4.16 &  $+$0.45 &  0.71 & 1.5271 &  1.0597 &   1.1424 & $-$4.20 \\
J054122.4-644254 &     5959 &  4.11 &  $+$0.32 &  0.61 & 1.8403 &  1.2300 &   1.3766 & $-$4.21 \\
J055244.5-404031 &     5812 &  4.33 &  $+$0.50 &  0.67 & 1.7876 &  1.1199 &   1.1734 & $-$4.20 \\
J121100.1-270721 &     5645 &  4.05 &  $+$0.33 &  0.70 & 1.6872 &  1.1024 &   1.1768 & $-$4.16 \\
J123710.3-380457 &     6158 &  4.14 &  $+$0.37 &  0.56 & 2.0516 &  1.3808 &   1.4507 & $-$4.22 \\
J173403.3-674016 &     5930 &  4.10 &  $+$0.36 &  0.62 & 1.9166 &  1.3119 &   1.3438 & $-$4.19 \\
J234144.8-544207 &     5664 &  4.11 &  $+$0.39 &  0.71 & 1.6673 &  1.1298 &   1.1380 & $-$4.17 \\
J042430.6-115901 &     5671 &  4.65 &  $+$0.31 &  0.70 & 1.7499 &  1.1887 &   1.2091 & $-$4.15 \\
J025410.6-383603 &     5509 &  4.18 &  $+$0.10 &  0.72 & 1.6805 &  1.1068 &   1.0652 & $-$4.13 \\
J233647.7-611842 &     5642 &  4.58 &  $+$0.31 &  0.71 & 1.7013 &  1.1827 &   1.1438 & $-$4.15 \\
J233510.7-221359 &     5564 &  4.50 &  $+$0.26 &  0.73 & 1.6943 &  1.1620 &   1.1674 & $-$4.13 \\
J072354.7-541313 &     6171 &  4.48 &  $+$0.36 &  0.56 & 2.0925 &  1.4589 &   1.4739 & $-$4.21 \\
J132226.2-251507 &     6031 &  4.58 &  $+$0.41 &  0.60 & 1.9854 &  1.3080 &   1.3300 & $-$4.21 \\
J150415.1-134338 &     6153 &  4.02 &  $+$0.37 &  0.56 & 2.2246 &  1.4307 &   1.4708 & $-$4.20 \\
J125627.5-364145 &     5694 &  4.27 &  $+$0.36 &  0.69 & 1.7067 &  1.1167 &   1.2150 & $-$4.17 \\
J162337.6-122109 &     5625 &  4.13 &  $+$0.33 &  0.71 & 1.6801 &  1.1000 &   1.1705 & $-$4.16 \\
J162227.1-020641 &     6249 &  4.07 &  $+$0.36 &  0.53 & 2.1967 &  1.5587 &   1.5894 & $-$4.21 \\
J115349.0-250448 &     6204 &  4.46 &  $+$0.21 &  0.53 & 2.3081 &  1.5516 &   1.6375 & $-$4.18 \\
J161932.2-031114 &     5566 &  4.05 &  $+$0.25 &  0.72 & 1.6980 &  1.1627 &   1.1991 & $-$4.13 \\
J164834.2-010021 &     6152 &  4.01 &  $+$0.32 &  0.55 & 2.2170 &  1.4302 &   1.5492 & $-$4.19 \\
J001512.8-591327 &     5991 &  4.09 &  $+$0.39 &  0.60 & 2.0199 &  1.3121 &   1.3091 & $-$4.20 \\
J125537.6-251221 &     5653 &  4.46 &  $+$0.32 &  0.70 & 1.6275 &  1.0834 &   1.2227 & $-$4.17 \\
J104750.6-094644 &     5682 &  4.03 &  $+$0.41 &  0.70 & 1.6954 &  1.1926 &   1.1499 & $-$4.17 \\
J120050.4-011600 &     5635 &  4.07 &  $+$0.30 &  0.70 & 1.5972 &  1.1465 &   1.2312 & $-$4.16 \\
J041226.0-424002 &     6186 &  4.08 &  $+$0.23 &  0.53 & 2.3249 &  1.5627 &   1.5944 & $-$4.18 \\
J205209.8-351619 &     5722 &  4.46 &  $+$0.48 &  0.70 & 1.6318 &  0.9633 &   1.1774 & $-$4.21 \\
J101034.1-183950 &     5971 &  4.04 &  $+$0.44 &  0.62 & 2.0012 &  1.3298 &   1.3752 & $-$4.19 \\
J220639.2-452731 &     5657 &  4.18 &  $+$0.38 &  0.71 & 1.7389 &  1.1812 &   1.2092 & $-$4.15 \\
J193108.0-620917 &     5957 &  4.08 &  $+$0.35 &  0.61 & 1.9658 &  1.3361 &   1.3841 & $-$4.18 \\
J045246.4-392443 &     5594 &  4.12 &  $+$0.34 &  0.72 & 1.6051 &  1.0651 &   1.0789 & $-$4.17 \\
J123112.5-290859 &     5627 &  4.37 &  $+$0.44 &  0.73 & 1.5212 &  1.0662 &   1.0946 & $-$4.19 \\
J134102.1-214222 &     5816 &  4.15 &  $-$0.20 &  0.59 & 1.9663 &  1.2521 &   1.4561 & $-$4.14 \\
J161951.5-101937 &     5749 &  4.05 &  $+$0.47 &  0.69 & 1.6880 &  1.1669 &   0.9966 & $-$4.21 \\
J100455.5-185152 &     6196 &  4.10 &  $+$0.37 &  0.55 & 2.1964 &  1.4748 &   1.5595 & $-$4.20 \\
J100455.5-185152 &     6212 &  4.03 &  $+$0.34 &  0.54 & 2.2815 &  1.4693 &   1.5793 & $-$4.20 \\
J154039.8-811212 &     5915 &  4.15 &  $+$0.40 &  0.63 & 1.9042 &  1.2596 &   1.2605 & $-$4.20 \\
J104628.7-212038 &     5637 &  4.03 &  $+$0.28 &  0.70 & 1.6855 &  1.1855 &   1.1335 & $-$4.16 \\
J202715.2-352857 &     5962 &  4.10 &  $+$0.36 &  0.61 & 1.8009 &  1.2226 &   1.2776 & $-$4.22 \\
J200958.0-833423 &     6039 &  4.04 &  $+$0.27 &  0.58 & 2.0520 &  1.3744 &   1.4294 & $-$4.19 \\
J160654.8-102640 &     5698 &  4.26 &  $+$0.32 &  0.69 & 1.6287 &  1.1429 &   1.2046 & $-$4.18 \\
J101704.0-183911 &     5584 &  4.34 &  $+$0.31 &  0.72 & 1.5667 &  1.1629 &   1.1592 & $-$4.15 \\
J200531.9-522958 &     6243 &  4.62 &  $+$0.42 &  0.55 & 2.1036 &  1.3833 &   1.5337 & $-$4.23 \\
J205020.8-632831 &     5661 &  4.52 &  $+$0.45 &  0.72 & 1.5948 &  1.1524 &   1.0569 & $-$4.18 \\
J005959.3-563005 &     6017 &  4.38 &  $+$0.46 &  0.61 & 1.9204 &  1.2974 &   1.3542 & $-$4.21 \\
J161308.0-023241 &     5664 &  4.04 &  $+$0.39 &  0.71 & 1.6680 &  1.1008 &   1.0988 & $-$4.18 \\
J002332.6-651126 &     5696 &  4.24 &  $+$0.37 &  0.69 & 1.6703 &  1.1759 &   1.2004 & $-$4.17 \\
J204427.1-215105 &     5581 &  4.30 &  $+$0.21 &  0.71 & 1.6031 &  1.1328 &   1.1780 & $-$4.15 \\
J222244.5-505907 &     6194 &  4.10 &  $+$0.30 &  0.54 & 2.2016 &  1.5159 &   1.5361 & $-$4.20 \\
J025546.1-520410 &     5565 &  4.20 &  $+$0.29 &  0.73 & 1.5535 &  1.1005 &   1.1405 & $-$4.16 \\
J020012.8-754531 &     6088 &  4.32 &  $+$0.39 &  0.58 & 1.9860 &  1.3693 &   1.4284 & $-$4.21 \\
J002115.0-302426 &     5547 &  4.04 &  $+$0.25 &  0.73 & 1.6367 &  1.1509 &   1.1836 & $-$4.13 \\
J024315.6-465558 &     5627 &  4.09 &  $+$0.29 &  0.71 & 1.6020 &  1.1070 &   1.2214 & $-$4.16 \\
J152528.8-231340 &     6079 &  4.05 &  $+$0.21 &  0.56 & 2.0952 &  1.4794 &   1.5291 & $-$4.18 \\
J211323.6-454007 &     5581 &  4.01 &  $+$0.35 &  0.73 & 1.6156 &  1.1018 &   1.0387 & $-$4.17 \\
J211929.7-124054 &     6217 &  4.52 &  $+$0.35 &  0.55 & 2.1974 &  1.5144 &   1.5186 & $-$4.21 \\
J233948.7-504249 &     5878 &  4.11 &  $+$0.36 &  0.63 & 1.8499 &  1.2668 &   1.3492 & $-$4.18 \\
J005259.3-225545 &     5636 &  4.32 &  $+$0.39 &  0.72 & 1.6536 &  1.1506 &   0.9505 & $-$4.18 \\
J050940.2-610616 &     6033 &  4.10 &  $+$0.39 &  0.59 & 2.0205 &  1.3912 &   1.3681 & $-$4.20 \\
J040854.2+000050 &     5641 &  4.02 &  $+$0.45 &  0.72 & 1.5615 &  1.0312 &   1.1727 & $-$4.18 \\
\hline
\end{tabular}
\end{small}
\tablefoot{Stellar effective temperature, surface gravity, and metallicity are from RAVE DR6, (B--V)$_0$ was converted from $T_{\rm eff}$. The last four columns are the (resolution) corrected fluxes for each of the IRT lines and the combined radiative loss in logarithmic form, respectively.}
\end{table*}

\begin{table*}
\caption{
Metallicity correction.
}\label{TAppMetal}
\begin{small}
\begin{tabular}{lcccccccccccc}
\hline\hline \noalign{\smallskip}
         &   \multicolumn{4}{c}{IRT-1}      & \multicolumn{4}{c}{IRT-2}  &  \multicolumn{4}{c}{IRT-3}\\
B--V bin & $\mathscr{F}_{\mathrm{Fe/H=0}}$ & $a$ & $b$ & $c$ & $\mathscr{F}_{\mathrm{Fe/H=0}}$ & $a$ & $b$ & $c$ & $\mathscr{F}_{\mathrm{Fe/H=0}}$ & $a$ & $b$ & $c$ \\
\noalign{\smallskip}\hline \noalign{\smallskip}
0.53--0.54 & 2.52 & $-$38.45 & 32.86 & $-$6.96  & 1.78 & $-$19.98 & 24.92 & $-$7.64     & 1.85 & $-$17.80 & 21.81 & $-$6.55  \\
0.54--0.55 & 2.49 & $-$27.41 & 24.27 & $-$5.31  & 1.76 & $-$26.95 & 33.33 & $-$10.18    & 1.82 & $-$19.59 & 24.07 & $-$7.27  \\
0.55--0.56 & 2.47 & $-$31.09 & 27.63 & $-$6.08  & 1.76 & $-$11.64 & 15.79 & $-$5.19     & 1.81 & $-$14.05 & 18.01 & $-$5.63  \\
0.56--0.57 & 2.40 & $-$41.26 & 36.81 & $-$8.15  & 1.72 & $-$7.93  & 12.23 & $-$4.39     & 1.77 & $-$15.56 & 20.40 & $-$6.51  \\
0.57--0.58 & 2.41 & $-$29.07 & 27.03 & $-$6.21  & 1.73 & $-$12.94 & 17.75 & $-$5.93     & 1.79 & $-$14.65 & 19.32 & $-$6.22  \\
0.58--0.59 & 2.41 & $-$20.55 & 20.14 & $-$4.84  & 1.74 & $-$15.33 & 21.05 & $-$7.06     & 1.79 & $-$12.44 & 17.14 & $-$5.73  \\
0.59--0.60 & 2.36 & $-$52.12 & 48.32 & $-$11.13 & 1.70 & $-$28.59 & 37.93 & $-$12.44    & 1.75 & $-$29.75 & 38.25 & $-$12.16 \\
0.60--0.61 & 2.27 & $-$29.20 & 28.52 & $-$6.88  & 1.64 & $-$16.92 & 23.71 & $-$8.14     & 1.68 & $-$22.36 & 29.77 & $-$9.76  \\ 
0.61--0.62 & 2.24 & $-$31.80 & 31.24 & $-$7.60  & 1.61 & $-$14.94 & 21.33 & $-$7.45     & 1.65 & $-$16.83 & 23.23 & $-$7.87  \\
0.62--0.63 & 2.20 & $-$26.89 & 27.36 & $-$6.87  & 1.59 & $-$15.71 & 22.75 & $-$8.07     & 1.63 & $-$11.99 & 17.71 & $-$6.35  \\
0.63--0.64 & 2.17 & $-$33.77 & 34.06 & $-$8.52  & 1.57 & $-$25.79 & 36.14 & $-$12.52    & 1.61 & $-$13.35 & 19.54 & $-$6.98  \\
0.64--0.65 & 2.16 & $-$29.25 & 30.22 & $-$7.72  & 1.57 & $-$14.66 & 22.10 & $-$8.14     & 1.60 & $-$16.68 & 24.34 & $-$8.70  \\
0.65--0.66 & 2.14 & $-$27.92 & 29.44 & $-$7.67  & 1.56 & $-$17.99 & 26.96 & $-$9.92     & 1.59 & $-$19.57 & 28.65 & $-$10.31 \\
0.66--0.67 & 2.10 & $-$26.59 & 28.73 & $-$7.67  & 1.53 & $-$9.84  & 16.16 & $-$6.40     & 1.56 & $-$17.78 & 26.71 & $-$9.85  \\
0.67--0.68 & 2.07 & $-$42.03 & 44.82 & $-$11.86 & 1.51 & $-$23.04 & 34.82 & $-$12.97    & 1.54 & $-$18.77 & 28.42 & $-$10.56 \\
0.68--0.69 & 2.01 & $-$23.74 & 26.49 & $-$7.29  & 1.47 & $-$18.79 & 29.20 & $-$11.16    & 1.49 & $-$14.73 & 22.99 & $-$8.78  \\
0.69--0.70 & 1.99 & $-$34.92 & 38.67 & $-$10.61 & 1.44 & $-$13.22 & 21.52 & $-$8.56     & 1.46 & $-$16.86 & 26.30 & $-$10.07 \\
0.70--0.71 & 1.93 & $-$36.42 & 41.32 & $-$11.61 & 1.40 & $-$12.23 & 21.00 & $-$8.77     & 1.43 & $-$18.38 & 29.45 & $-$11.59 \\
0.71--0.72 & 1.92 & $-$34.10 & 39.41 & $-$11.28 & 1.40 & $-$6.21  & 12.17 & $-$5.55     & 1.42 & $-$17.08 & 28.03 & $-$11.27 \\
0.72--0.73 & 1.88 & $-$33.59 & 39.41 & $-$11.45 & 1.36 & $-$15.99 & 27.20 & $-$11.33    & 1.38 & $-$21.07 & 34.18 & $-$13.65 \\
\noalign{\smallskip}\hline
\end{tabular}
\end{small}
\tablefoot{$\mathscr{F}_{\mathrm{Fe/H=0}}$ is the flux at zero metallicity in $10^6$ \ecms, and $a$, $b$, and $c$ are the coefficients for a second order polynomial fit in form $a + bx + cx^2$, where $x$ is the $\mathscr{F}_{\mathrm{Fe/H}}$ in $10^6$ \ecms.}
\end{table*}

\end{appendix}

\end{document}